\pdfoutput=1

\documentclass[11pt]{article}

\usepackage[]{EMNLP2022}

\usepackage{times}
\usepackage{latexsym}

\usepackage[T1]{fontenc}

\usepackage[utf8]{inputenc}

\usepackage{microtype}

\usepackage{inconsolata}

\usepackage{graphicx}
\usepackage{booktabs}
\usepackage{multirow}
\usepackage{pifont}
\usepackage{float}
\usepackage{siunitx}
\usepackage{amssymb}
\usepackage{subfig}
\usepackage{arydshln}
\usepackage{numprint}
\usepackage[dvipsnames]{xcolor}
\usepackage{tabularx}
\usepackage{enumitem}

\newcommand{\cmark}{\ding{51}}%
\newcommand{\xmark}{\ding{55}}%
\newcolumntype{Y}{>{\centering\arraybackslash}X}

\title{Incorporating Relevance Feedback for Information-Seeking Retrieval \\ using Few-Shot Document Re-Ranking}
\author{
Tim Baumgärtner,$^{1}$ 
Leonardo F. R. Ribeiro,$^{1*}$ 
Nils Reimers,$^{2}$ 
Iryna Gurevych$^{1}$ \\
$^{1}$Ubiquitous Knowledge Processing Lab (UKP Lab), \\ Department of Computer Science and Hessian Center for AI (hessian.AI), \\Technical University of Darmstadt \\
$^{2}$cohere.ai \\
\url{www.ukp.tu-darmstadt.de}
}

\begin{document}
\maketitle
\def\thefootnote{*}\footnotetext{Now affiliated with Amazon Alexa AI.}
\def\thefootnote{\arabic{footnote}}
\begin{abstract}
Pairing a lexical retriever with a neural re-ranking model has set state-of-the-art performance on large-scale information retrieval datasets. This pipeline covers scenarios like question answering or navigational queries, however, for information-seeking scenarios, users often provide information on whether a document is relevant to their query in form of clicks or explicit feedback. Therefore, in this work, we explore how relevance feedback can be directly integrated into neural re-ranking models by adopting few-shot and parameter-efficient learning techniques. Specifically, we introduce a kNN approach that re-ranks documents based on their similarity with the query and the documents the user considers relevant. Further, we explore Cross-Encoder models that we pre-train using meta-learning and subsequently fine-tune for each query, training only on the feedback documents. To evaluate our different integration strategies, we transform four existing information retrieval datasets into the relevance feedback scenario. Extensive experiments demonstrate that integrating relevance feedback directly in neural re-ranking models improves their performance, and fusing lexical ranking with our best performing neural re-ranker outperforms all other methods by $5.2\%$ nDCG@20.\footnote{The code is available at \url{https://github.com/UKPLab/incorporating-relevance}}
\end{abstract}

\section{Introduction}
User queries can be categorized as navigational (retrieving a specific document), transactional (retrieving a website to perform a particular action) or informational \citep{broder_taxonomy}. For information-seeking queries, users might want to learn about a new topic or might be unfamiliar with the search domain. Therefore they potentially do not use common keywords of the domain which decreases performance \citep{furnas-1987-vocab}. Furthermore, they might want to find complementary information from diverse sources or consider different aspects of a topic \citep{clarke-2008-novelty-diversity}. Lastly, information-seeking queries can also be used to keep up with the latest developments on a topic.

Concretely, these queries are encountered during scientific literature review \citep{voorhees2021trec, dasigi-etal-2021-dataset}, when looking for news and background information \citep{soboroff2018trec}, during argument retrieval, \citep{bondarenko2020overview} or in the legal context for case law retrieval \citep{locke-zuccon-test-2018}. 

Formulating effective queries to satisfy the complex information need in these scenarios is difficult. On the contrary, a user can easily judge whether a document is relevant to their query. Therefore, information obtained from the user when interacting with the search results, known as \emph{relevance feedback}, can be used in the search. This can be obtained implicitly from click logs \citep{joachims-2002-clickthrough} or explicitly by asking users whether a document is relevant \citep{rocchio1971relevance}. We focus on explicit feedback because it is clean compared to implicit feedback and existing information retrieval datasets can be transformed into this scenario.
In both settings, the amount of feedback is limited, since users will provide feedback only on a few documents.

Incorporating relevance feedback in information retrieval (IR) systems is well-established for lexical retrieval \citep{rocchio1971relevance, 10.1145/383952.383972, zhai2001model}. These systems incorporate the feedback by expanding the query with terms extracted from relevance feedback documents. While these approaches can alleviate the lexical gap, they inherently struggle with semantics because they represent text as a bag of words. Additionally, lexical query expansion methods have the disadvantage that their latency increases with the number of query terms \citep{wu2013incremental}.

To mitigate these issues, neural retrieval and re-ranking methods have been proposed and recently outperformed lexical retrieval \citep{gillick-etal-2019-learning, nogueira-cho-2019-reranking, karpukhin-etal-2020-dense, khattab-zaharia-2020-colbert}. State-of-the-art retrieval results are obtained in a two-stage setup: First, an efficient and recall-optimized retrieval method (e.g. dense or lexical retrieval) retrieves an initial set of documents. Subsequently, a neural re-ranker optimizes the rank of the documents. However, there exists no neural re-ranking model that directly incorporates relevance feedback. 

To this end, we explore how relevance feedback can directly be integrated into neural re-ranking models. This is difficult because state-of-the-art models have millions of parameters and require a large amount of training data, while only a limited amount of relevance feedback per query is available. We make use of recent advances in parameter-efficient fine-tuning \citep{pmlr-v97-houlsby19a, ben-zaken-etal-2022-bitfit} and few-shot learning \citep{NIPS2017_cb8da676, pmlr-v70-finn17a} to address the challenges of model re-usability and learning from limited data. Concretely, we present a kNN approach that re-ranks documents based on their similarity to the feedback documents. We further propose to fine-tune a re-ranking model from only the relevance feedback for each query. We explore the effectiveness of our approach with a varying number of feedback documents and evaluate its computational efficiency. To evaluate our models, we transform four existing IR datasets into the re-ranking with relevance feedback setup. Our final model combines the strengths of lexical and neural re-ranking using reciprocal rank fusion \citep{cromack-2009-reciprocal-rank-fusion}.

In summary, our contributions are as follows:
\begin{itemize}
    \item We propose a few-shot learning task for information retrieval, specifically adopting the two-stage retrieve and re-rank settings to incorporate relevance feedback, both in the retrieval as well as in the re-ranking.
    \item We outline retrieval scenarios for the task and how to transform existing IR datasets into the few-shot retrieve and re-rank setup.
    \item We present novel re-ranking methods that directly incorporate relevance feedback leveraging few-shot learning and parameter-efficient techniques. We evaluate their efficiency and demonstrate their effectiveness through extensive experiments and across different datasets.
\end{itemize}

\section{Related Work}

\begin{table*}[t]
\small
\centering
\resizebox{\linewidth}{!}{
\begin{tabular}{@{}llccccc@{}}

\toprule
Dataset & Domain & Docs & Doc. Length & Queries & Q. Length & Judgments \\ \midrule
Robust04 \citep{voorhees2004overview} & News & 528k & 476.40 & 148 & 16.76 & 1287.14 (\textpm{501}) \\
TREC-Covid \citep{voorhees2021trec} & Biomedical & 191k & 158.87 & 50 & 10.96 & 1370.36 (\textpm{323}) \\
TREC-News \citep{soboroff2018trec} & News & 595k & 686.65 & 34 & 12.03 & 258.85 (\textpm{82}) \\
Webis-Touché \citep{bondarenko2020overview} & Debates & 383k & 289.34 & 49 & 6.67 & 49.76 (\textpm{7}) \\ \bottomrule

\end{tabular}}
\caption{Datasets used for the few-shot re-ranking task. Length: average number of words. Judgments: average number (and standard deviation) of relevant and non-relevant judged documents per query. The datasets have been filtered to only include queries with a minimum number of relevant and non-relevant documents.}
\label{tab:datasets}
\vspace{-2mm}
\end{table*}

\subsection{Information Retrieval Approaches}
Traditionally, lexical approaches have been used for IR, such as TF-IDF and BM25 \citep{robertson-zaragoza-2009-bm25}. However, these systems cannot model lexical-semantic relations between query and document (the document and query are treated as bag of words) and suffer from the lexical gap \citep{10.1145/345508.345576}, e.g., when synonyms are used.

Recently, dense retrieval methods have shown promising results, outperforming lexical approaches \citep{gillick-etal-2019-learning, karpukhin-etal-2020-dense, khattab-zaharia-2020-colbert}. Contrary to lexical systems, they can discover semantic matches between a query and a document, thereby overcoming the lexical gap. Dense retrieval methods learn query and document representations in a shared, high-dimensional space. This is enabled by large-scale pre-training \citep{devlin-etal-2019-bert} and training on IR datasets of considerable size \citep{nguyen2016ms, kwiatkowski-etal-2019-natural}. 
After training, the model computes a document index holding a representation for each document in the corpus. At inference, a query representation is compared to each document vector using maximum inner product search \citep{johnson2019billion}. 

However, applying dense retrieval to our setup is not practical. We aim to fine-tune the model for every query, therefore, the precomputed document index would become out of sync with the model and might not yield optimal results \citep{realm}. Since the document index is very large, re-encoding it would create an unreasonable computational overhead. Thus, we do not experiment with dense retrieval models that rely on a precomputed document index.

Similar to dense retrieval, neural re-ranking models have profited from pre-training and training on large datasets. The predominant approach is to use a Cross-Encoder (CE) model that takes both query and document as input to directly compute a relevance score. Contrary to dense retrieval models, this enables direct query-document interactions. Since this approach does not allow to pre-compute representations and is compute-intensive, it is generally paired with a more efficient first-stage retrieval method (dense or lexical) and subsequently applied to the top retrieved documents. Particularly combined with lexical retrieval methods, neural re-ranking yields state-of-the-art performance \citep{thakur2021beir}.

\subsection{Relevance Feedback}
Relevance feedback has mostly been integrated into IR systems by modifying the query using the feedback documents and subsequently performing a second round of retrieval. \citet{rocchio1971relevance} propose a linear combination of the vectors of the query, the relevant and non-relevant feedback documents to obtain a new query vector, which is more similar to the relevant documents. Another approach is to use language models of the query and documents to obtain new terms \citep{10.1145/383952.383972, zhai2001model}. Recently, \citet{shahrzad-etal-2021-ceqe} use the similarity between contextualized query and document word embeddings to extract terms for query expansion. Similarly, \citet{zheng-etal-2020-bert} use BERT to obtain document chunks for expansion and subsequently compute the relevancy by summing over chunk-document relevance. While these works leverage advances in pre-trained language modeling for selecting query terms, they eventually rely only on lexical retrieval, potentially missing semantic matches in the document collection. While we also use lexical retrieval with query expansion for the second stage, we additionally update a re-ranking model based on the relevance feedback and employ it on the second stage retrieval results.

Other works directly incorporate relevance feedback into neural retrieval. \citet{ai-2018-learning} train a model that sequentially encodes the top document representations from the first stage retrieval. The documents are subsequently re-ranked using an attention mechanism between the final and intermediate representations of the model. \citet{yu-2021-improving} further fine-tune the query encoder of a dense retrieval model to additionally take the top documents from a first retrieval stage as input. While these works directly incorporate first-stage retrieval documents into their model, they require large annotated datasets to train their models. Furthermore, adding the feedback documents to the input is sub-optimal due to large memory requirements of transformer models with growing input size. Our approach overcomes this by using the relevance feedback to update the model parameters instead of providing it as input.

Most similar to our work, \citet{lin2019simplest} propose to learn a re-ranker using machine learning classifiers (logistic regression and support vector machines) based on lexical features from the top and bottom retrieved documents. They show that this simple approach improves over query expansion and neural approaches like NPRF \citep{li-etal-2018-nprf}. In contrast to our work, they use pseudo-relevance feedback and simple classification approaches as a re-ranking model. Moreover, we use explicit relevance feedback since the automatic selection of non-relevant documents is challenging. Depending on the query and document collection the number of relevant documents varies significantly. For one query there might only be few relevant documents in which case irrelevant documents could be selected from higher ranks. Another query might return a large set of relevant documents in which case pseudo-irrelevant documents would actually need to be selected from lower ranks. User feedback on the other hand does not have this disadvantage. However, since users will only give feedback on limited documents, the models used by \citet{lin2019simplest} cannot be trained from explicit feedback. Therefore, we opt for few-shot learning combined with pre-trained re-ranking models. Furthermore, the user-selected documents also provide a form of interpretability to the re-ranking model.

In summary, using pseudo and explicit relevance feedback in lexical models via query expansion has shown to improve retrieval performance. Furthermore, neural retrieval and re-ranking models have shown promising results, outperforming lexical methods. While there exists related work that combines neural models with query expansion, they are applied to pseudo-relevance feedback and use state-of-the-art models only for determining query expansion terms. Other methods are limited in the amount of feedback and require large training datasets for fine-tuning. In this work, we leverage few-shot learning techniques to directly update a re-ranking model based on explicit feedback.

\section{Datasets}
Large-scale IR datasets mostly target use cases where a user has a less complex information need, e.g., looking for a factoid answer \citep{nguyen2016ms, kwiatkowski-etal-2019-natural, zhang-etal-2021-mr}. These are usually sparsely annotated, i.e. there is only a single (or few) judged relevant documents per query. However, for our information-seeking use case, we are interested in queries where many relevant documents exist. Therefore, we select datasets where a large set of relevant documents per query are judged. Further, the datasets should target suitable use cases containing queries that have a diverse set of relevant documents. For example, the query ``What is the origin of Covid-19'' from TREC-Covid, has relevant documents about the geographical location of the first cases, the genetic origins of the virus, and animals that likely have transmitted the disease to humans.

Specifically, we consider Robust04 \citep{voorhees2004overview}, TREC-Covid \citep{voorhees2021trec}, TREC-News \citep{soboroff2018trec}, and Webis-Touché \citep{bondarenko2020overview}. An overview of all datasets with their statistics is provided in Table~\ref{tab:datasets}.\footnote{More details on the datasets in Appendix \ref{sec:appendix-datasets}.} We transform these datasets into the few-shot re-ranking setup by including only queries with at least $32$ judged relevant and non-relevant documents in the BM25 top 1000 results with the query. Any queries with fewer judged documents are discarded because they provide little evaluation power, because we remove the feedback documents from the evaluation. Moreover, this filter ensures that enough judged documents are present for a robust evaluation. 

For our experiments, we create training, validation and test splits in a 3:1:1 ratio, by randomly assigning each query to one set. We further conduct three random shuffles over the assignment of a query into the training, validation, and test set. We report the averaged results over the shuffles. 

\section{Task Setup}\label{sec:task-setup}
\begin{figure}[t]
\centering
\includegraphics[width=1\linewidth]{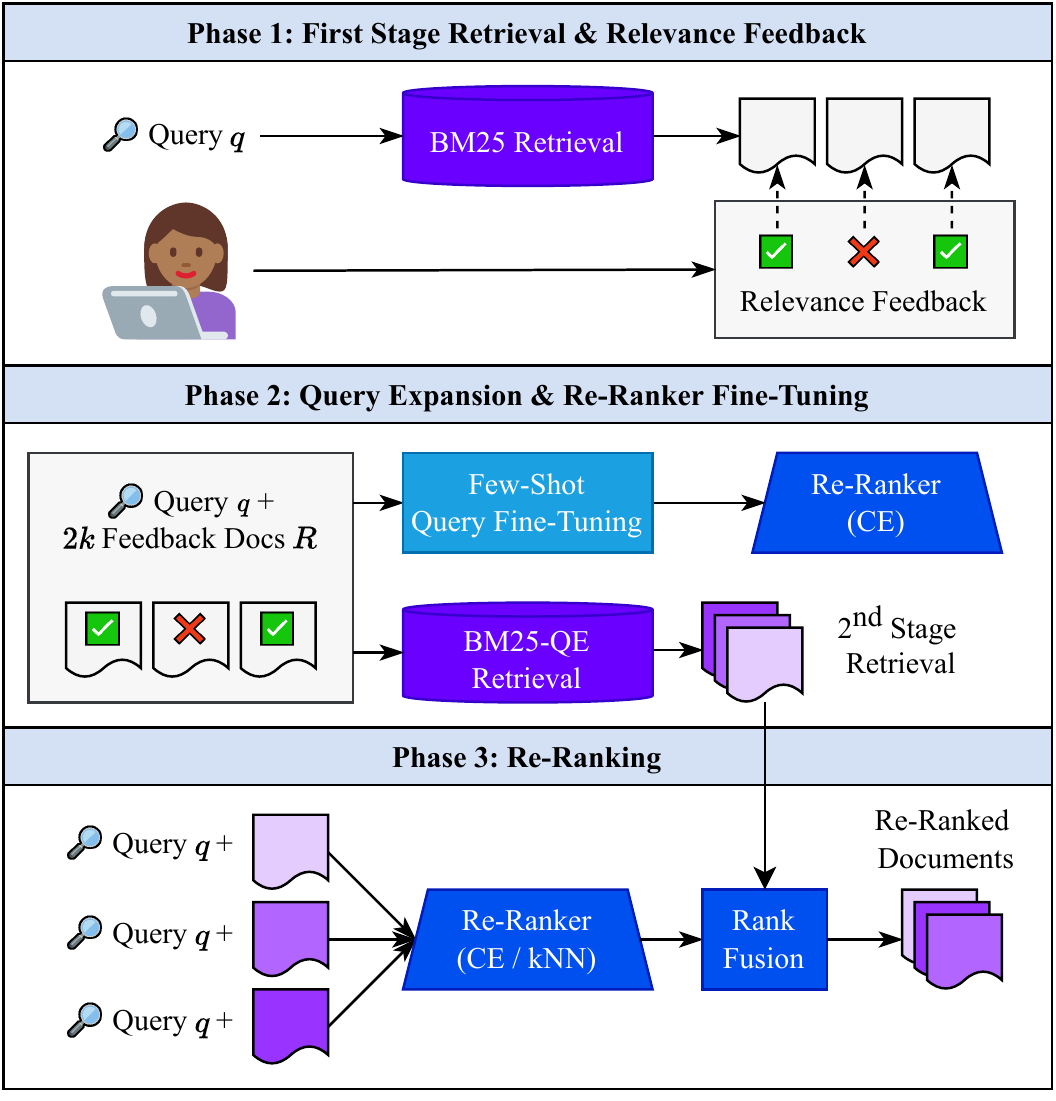}
\caption{The three phases of our proposed few-shot retrieve and re-rank setup. \textbf{Phase 1}: Documents are retrieved using the query $q$, and relevance feedback is obtained from a user. \textbf{Phase 2}: The query $q$ and feedback documents $R$ are used for query expansion and the second round of retrieval. Further, a re-ranking model is fine-tuned using the user-selected feedback documents. \textbf{Phase 3}: The documents are re-ranked using the fine-tuned re-ranker, obtaining the final document ordering. To improve performance, the ranking from the re-ranker and the second phase are be fused.} 
\label{fig:flow}
\end{figure}

To incorporate relevance feedback in any retrieval process, a multi-stage approach is required. We propose a multi-phase task setup which is visualized in Figure~\ref{fig:flow}. In \textbf{Phase 1}, the relevance feedback is collected from the user after a first retrieval. The selected documents refine the information need and provide additional insight into what is relevant to the query. In \textbf{Phase 2}, the feedback is processed and a second retrieval is conducted while the re-ranking model is trained on the selected feedback documents. This phase returns documents that are more relevant to the user's information need. Ultimately, in \textbf{Phase 3}, the documents obtained previously are re-ranked based on the tailored re-ranker.

Specifically, in \textbf{Phase 1}, an initial retrieval is conducted with the query $q$ against the document collection. For lexical retrieval, we use BM25 as it is robust in a zero-shot setting on a diverse set of domains \citep{thakur2021beir}. Next, we select the top $k \in \{2, 4, 8\}$ relevant and non-relevant documents from the first-stage retrieval according to the judgments in the dataset, i.e., there are 2$k$ documents selected per query.

We refer to these documents as \textit{feedback documents} $R$. By selecting the top judged and retrieved documents, this process simulates a user providing relevance feedback.\footnote{We also experimented with selecting judged documents randomly but preliminary experiments showed that this generally leads to worse performance.} For our evaluation, we remove the feedback documents from the relevance judgments (i.e. we use the \textit{residual collection}, (\citeauthor{salton1990improving}, \citeyear{salton1990improving}) in order to evaluate the ability of the models to rank non-selected documents higher.

In \textbf{Phase 2}, we use the $2k$ feedback documents for query expansion and a second retrieval step. We extract $e$ terms per relevant feedback document and append them to the query resulting in the expanded query for second-stage retrieval.\footnote{We also experimented with negatively weighing terms from non-relevant documents. However, we find that this generally hurts performance.} Furthermore, the feedback documents are used to fine-tune a re-ranking model. Starting from a common base model, a new model is fine-tuned for every query. To exploit the small number of feedback documents most effectively, we employ few-shot learning when fine-tuning the re-ranker.

Finally, in \textbf{Phase 3}, the documents are scored using the query-specific re-ranker from the second phase. Additionally, the ranking from BM25-QE can be incorporated in the final document ranking. We experiment with different models and settings, details are described in \S\ref{sec:methods}.

The re-ranking could also be performed on the documents from the first-stage retrieval. However, since the feedback documents are available and query expansion generally improves recall (which is important for the re-ranking performance), we chose to not experiment with re-ranking the first-stage retrieval documents. This also improves the evaluation, because all models re-rank the same set of documents. 

\paragraph{Evaluation Metrics.} To measure the ranking performance we use nDCG@20 \citep{jarvelin2000ir} implemented by \textsc{pytrec\_eval} \citep{VanGysel2018pytreceval}. This metric considers graded relevance labels. We chose the cut-off at 20 to take the large number of relevant documents per query into account. Beyond ranking performance, we also focus on retrieval/re-ranking latency and parameter efficiency. The response time of IR systems is generally crucial for user satisfaction \citep{schurman2009performance}. Therefore, we evaluate the time for retrieval, query expansion, fine-tuning, and re-ranking. Since we fine-tune a model per query, the memory footprint of the model should be small. This allows keeping many models in memory at the same time or quickly reloading a model whenever a user revisits a query.

\section{Methods}\label{sec:methods}
\subsection{BM25 Query Expansion}
For the second-stage retrieval, we expand the original query $q$ with terms $e$ obtained from the relevant feedback documents. We experiment with a varying number of expansion terms $e \in \{4, 8, 16, 32, 64\}$ and also use all terms in the document for expansion which we refer to as \textit{all}. We do so by using Elasticsearch's MoreLikeThis feature,\footnote{\href{https://www.elastic.co/guide/en/elasticsearch/reference/current/query-dsl-mlt-query.html}{Elasticsearch: MoreLikeThis}} which extracts terms according to their TF-IDF score. For retrieval, the query and the extracted terms are combined, and the documents are scored according to BM25. This setup follows the query expansion technique described in \citet{rocchio1971relevance}. The ranking produced by BM25 query expansion (BM25-QE) serves as the lexical baseline in our experiments. 

\subsection{Re-ranker}
In this section, we detail the different approaches employed for document re-ranking: kNN, Cross-Encoder, and Rank Fusion. 

\subsubsection{kNN}
The kNN approach is based on a dense retrieval model that computes a high-dimensional, semantic text representation. Specifically, we use the transformer-based MiniLM \citep{NEURIPS2020_3f5ee243} model that was fine-tuned on a diverse set of training datasets.\footnote{\href{https://discuss.huggingface.co/t/train-the-best-sentence-embedding-model-ever-with-1b-training-pairs/7354}{https://discuss.huggingface.co/t/train-the-best-sentence-embedding-model-ever-with-1b-training-pairs/7354}} 

We use the 6-Layer model since its counterpart with 12 layers only provides marginally better performance albeit requiring twice the compute.

To obtain a document score $s_i$, we compute the similarity between the query $q$ and the document $d_i$ and add the sum of similarities between the relevant feedback documents $d_j \in R^{+}$ and $d_i$. We use cosine-similarity as similarity function $f$. This is expressed in Equation~\ref{eq:knn}. 
\begin{equation}\label{eq:knn}
s_i = f(d_i, q) + \sum_{d_j \in R^{+}} f(d_{i}, d_{j})
\end{equation}
The kNN setup resembles Prototypical Networks \citep{NIPS2017_cb8da676}, however, instead of having a single, averaged point in vector space representing a class, we have $k+1$ points (all relevant feedback documents and the query). In this setting, the model weights are not updated, instead, we use the document and query encodings for finding similar documents.

\subsubsection{Cross-Encoder (CE)}
For re-ranking with a Cross-Encoder, we employ the 6-Layer MiniLM model fine-tuned on MS MARCO \citep{hofstatter2020improving}. We experiment with zero-shot, query fine-tuning, and meta-learning approaches.

\paragraph{Zero-Shot.}As a baseline, we do not perform any fine-tuning and re-rank the documents with the pre-trained model. Zero-shot only refers to not fine-tuning the re-ranking model, however, we still re-rank the documents obtained with query expansion for comparability reasons.
\paragraph{Query Fine-Tuning.} We update the re-ranker using few-shot supervised learning with the $2k$ feedback documents. We optimize the Binary Cross-Entropy and use the validation set to determine the learning rate and the number of training steps that perform best on average according to the nDCG@20 score. We refer to this as \textit{CE Query-FT}.

\paragraph{Meta-Learning.}In order to optimize the model for quick adaption to new queries, we also explore using model-agnostic meta-learning (MAML) \citep{pmlr-v70-finn17a}. Meta-learning is generally defined over a set of tasks (as opposed to a set of training samples). Therefore, we treat each query with the respective feedback documents as its own task. This is reasonable since we model the relevance of a document in the context of the query.
The training process consists of two stages: (1) First, the model $g$ is optimized on the training dataset. Each batch consists of two tasks $T_1$ and $T_2$, each comprising a query and the respective $2k$ feedback documents. The model parameters $\theta$ are updated using $T_1$ optimizing the Binary Cross-Entropy on the feedback documents with learning rate $\alpha$. We obtain new parameters $\theta'$ from this step. We show this formally in Equation~\ref{eq:maml-1} for a single step.
\begin{equation}\label{eq:maml-1}
    \theta' = \theta - \alpha \nabla_{\theta} \mathcal{L}(g_\theta ; T_1)
\end{equation}

Subsequently, the new parameters are evaluated on their ability to adapt to the second task $T_2$ by computing the loss of the predictions made by the model $g_{\theta'}$. By backpropagating through this entire process (i.e. computing the gradients w.r.t. to $\theta$), the original parameters of the model are optimized: 
\begin{equation}\label{eq:maml-2}
    \theta'' = \theta - \alpha \nabla_{\theta} \mathcal{L}(g_{\theta'} ; T_2)
\end{equation}

Intuitively, the loss in Equation~\ref{eq:maml-2} will be low, if the parameters $\theta'$ can quickly adapt to $T_2$. We refer the reader to \citet{pmlr-v70-finn17a} for a more detailed overview of the training process using MAML. In our training process, we only use a single task, i.e. one query with its respective feedback documents, per step due to the limited amount of training data. We find the hyperparameters according to the zero-shot performance on the validation dataset. (2) Once the MAML training concludes, the model is updated per query as detailed in the Query Fine-Tuning paragraph. We call this method \textit{CE MAML + Query FT}.

\paragraph{Parameter Efficiency.}For all Cross-Encoder methods we only update the bias layers as proposed by \citet{ben-zaken-etal-2022-bitfit}. This keeps the number of tunable parameters and the memory footprint of the models very small. Using this method only $0.11\%$ of the parameters are updated. Compared to adapters \citep{pmlr-v97-houlsby19a}, tuning the biases is advantageous because the parameters are already tuned and not randomly initialized.

\subsubsection{Rank Fusion}
We also investigate merging the rankings produced by BM25-QE and the neural re-ranking model using Reciprocal Rank Fusion (RRF) \citep{cromack-2009-reciprocal-rank-fusion}. The final ranking is computed according to Equation \ref{eq:rank-fusion}, where $s_i$ is the fused score of document $d_i$, $h$ is the ranking function returning the rank of a document and $c$ is a constant decreasing the impact of the top-scored documents.\footnote{We leave the constant at the default value of 60.}

\begin{equation}\label{eq:rank-fusion}
    s_i = \sum_{h \in H} \frac{1}{c+h(d_i)}
\end{equation}
This approach has the advantage of being agnostic to the relevance scores assigned to the documents by the models because it only uses their rankings. Using the relevance scores directly is problematic when the scores of the models are in different ranges.\footnote{E.g., BM25 can produce large scores per document as it is a sum of scores, while binary classification models like Cross-Encoder models produce scores between 0-1.} Intuitively, RRF can leverage diverse rankings to improve the result. Furthermore, it will rank documents higher that are strongly preferred by one ranking model than documents that are weakly preferred by multiple models.\footnote{For example, if $h_0(d_0) = 5$, $h_1(d_0) = 15$, $h_0(d_1)=10$ and $h_1(d_1)=10$ then $s_0 > s_1$.}

\section{Results}
\subsection{2\textsuperscript{nd} Stage Retrieval: Query Expansion}
We report recall@1000 results of the second stage retrieval with varying number of expansion terms~$e$ in Table~\ref{tab:query-expansion}. Note that by increasing $e$ the performance increases, reaching a maximum at $e = 8$. However, when further increasing $e$, the recall drops. We observe qualitatively that extracting more terms per document also includes more non-specific terms or even stop words which hurt performance. Based on the recall@1000 performance on the validation set (see Appendix \ref{sec:bm25-expansion}) we use the documents obtained by extracting $e = 16$ terms for the final re-ranking step. In this work, we do not focus on the first stage retrieval. For completeness, we report the results in Appendix~\ref{sec:first-stage-retrieval-results}.

\begin{table}
\small
\centering
\begin{tabularx}{\columnwidth}{@{}lYYYY@{}}
\toprule
$e$ & $k=2$ & $k=4$ & $k=8$ & Avg. \\ \midrule
4 & 0.6187 & 0.6300 & 0.6566 & 0.6351 \\
8 & \textbf{0.6280} & \textbf{0.6414} & \underline{0.6721} & \textbf{0.6472} \\
16 & \underline{0.6195} & \underline{0.6400} & \textbf{0.6736} & \underline{0.6444} \\
32 & 0.6039 & 0.6209 & 0.6477 & 0.6242 \\
64 & 0.5597 & 0.5729 & 0.5843 & 0.5723 \\
all & 0.5723 & 0.5771 & 0.5828 & 0.5774 \\ \bottomrule
\end{tabularx}
\caption{Recall@1000 results on the test set with varying number of expansion terms $e$ from each relevant document. Results are averaged over the shuffles.}
\label{tab:query-expansion}
\end{table}

\begin{table}[ht!]
\small
\centering
\resizebox{\columnwidth}{!}{

\begin{tabularx}{\columnwidth}{@{}lYYYYY@{}}
    \toprule
     & Robust & Covid & News & Touché & Avg. \\ \midrule
    \multicolumn{6}{l}{\textit{BM25-QE}} \\ \midrule
    $k=2$ & 0.4480 & 0.5632 & \underline{0.3846} & \underline{0.2602} & 0.4140 \\
    $k=4$ & 0.4843 & 0.6079 & 0.3877 & \underline{0.2558} & 0.4339 \\
    $k=8$ & 0.5568 & 0.6606 & 0.4049 & \textbf{0.2982} & 0.4801 \\ \cdashline{1-6}
    Avg. & 0.4964 & 0.6106 & 0.3924 & \underline{0.2714} & 0.4427 \\ \midrule
    \multicolumn{6}{l}{\textit{kNN}} \\ \midrule
    $k=2$ & 0.4259 & 0.6736 & 0.3492 & 0.1646 & 0.4033 \\
    $k=4$ & 0.4342 & 0.6789 & 0.3539 & 0.1697 & 0.4092 \\
    $k=8$ & 0.4698 & 0.7069 & 0.3925 & 0.1904 & 0.4399 \\ \cdashline{1-6}
    Avg. & 0.4433 & 0.6865 & 0.3652 & 0.1749 & 0.4175 \\ \midrule
    
    \multicolumn{6}{l}{CE Zero-Shot} \\ \midrule
    $k=2$ & 0.3937 & 0.6917 & 0.2955 & 0.1731 & 0.3885 \\
    $k=4$ & 0.4185 & 0.7018 & 0.3189 & 0.1767 & 0.4040 \\
    $k=8$ & 0.4335 & 0.7150 & 0.3285 & 0.1799 & 0.4142 \\ \cdashline{1-6}
    Avg. & 0.4152 & 0.7028 & 0.3143 & 0.1766 & 0.4022 \\ \midrule
    
    \multicolumn{6}{l}{\textit{CE Query-FT}} \\ \midrule
    $k=2$ & 0.4375 & 0.6833 & 0.2942 & 0.1887 & 0.4009 \\
    $k=4$ & 0.4786 & 0.7182 & 0.3463 & 0.2080 & 0.4378 \\
    $k=8$ & 0.5376 & \textbf{0.7677} & 0.3645 & 0.1975 & 0.4668 \\ \cdashline{1-6}
    Avg. & 0.4846 & 0.7231 & 0.3350 & 0.1981 & 0.4352 \\ \midrule
    \multicolumn{6}{l}{\textit{CE MAML + Query FT}} \\ \midrule
    $k=2$ & 0.4529 & \underline{0.7129} & 0.2526 & 0.2212 & 0.4099 \\
    $k=4$ & \underline{0.5079} & \textbf{0.7498} & 0.3358 & 0.2292 & 0.4557 \\
    $k=8$ & 0.5572 & 0.7449 & 0.3557 & 0.2201 & 0.4695 \\ \cdashline{1-6}
    Avg. & 0.5060 & \underline{0.7359} & 0.3147 & 0.2235 & 0.4450 \\ \midrule
    
    \multicolumn{6}{l}{\textit{Rank Fusion: kNN \& BM25-QE}} \\ \midrule
    $k=2$ & \underline{0.4635} & 0.6903 & 0.3783 & 0.2263 & \underline{0.4396} \\
    $k=4$ & 0.5020 & 0.6858 & \textbf{0.4228} & 0.2438 & \underline{0.4636} \\
    $k=8$ & \underline{0.5574} & 0.7470 & \textbf{0.4359} & 0.2744 & \underline{0.5037} \\ \cdashline{1-6}
    Avg. & \underline{0.5076} & 0.7077 & \textbf{0.4123} & 0.2482 & \underline{0.4689} \\ \midrule
    
    \multicolumn{6}{l}{\textit{Rank Fusion: CE MAML + Query FT \& BM25-QE}} \\ \midrule
    $k=2$ & \textbf{0.5164} & \textbf{0.7269} & \textbf{0.3934} & \textbf{0.2670} & \textbf{0.4759} \\
    $k=4$ & \textbf{0.5576} & \underline{0.7449} & \underline{0.4084} & \textbf{0.2701} & \textbf{0.4953} \\
    $k=8$ & \textbf{0.6380} & \underline{0.7489} & \underline{0.4148} & \underline{0.2809} & \textbf{0.5207} \\ \cdashline{1-6}
    Avg. & \textbf{0.5707} & \textbf{0.7402} & \underline{0.4055} & \textbf{0.2727} & \textbf{0.4973} \\ \bottomrule
    \end{tabularx}
}
\caption{nDCG@20 test set results averaged over three seeds with a varying number of feedback documents ($k$). In bold, the best performing model, the runner-up is underlined.}
\label{tab:main-results-ndcg20}
\vspace{-1mm}
\end{table}

\subsection{Re-Ranking Performance}
We report the nDCG@20 ranking performance in Table~\ref{tab:main-results-ndcg20} and additional zero-shot baselines in Appendix~\ref{sec:zero-shot-baselines}. We first note that increasing the amount of relevance feedback $k$ generally improves performance. 
Furthermore, we observe that BM25-QE already performs well. Neither the kNN approach, nor the Cross-Encoder zero-shot and Query FT, nor the wide variety of zero-shot models are able to outperform BM25-QE, except on TREC-Covid. We note a superior performance on Webis-Touché, although this task is the most challenging for neural models in our test suite. This agrees with related work that indicates that BM25 beats all other methods on this task \citep{thakur2021beir}. 
When looking at the CE experiments, we observe incremental performance increases when the relevance feedback is integrated. CE zero-shot is outperformed by query fine-tuning, which is subsequently outperformed when MAML training is added. This shows that our proposed direct integration of relevance feedback in the model is effective and that the parameters obtained by MAML training are better able to adapt to new queries given the relevance feedback. This method also slightly outperforms BM25-QE.

Finally, combining the rankings of the lexical retrieval and neural re-ranker is particularly effective. While different methods excel at each dataset (e.g. BM25 on Webis-Touché or neural models on TREC-Covid), the rank fusion is able to mitigate the weaknesses of one model successfully. Moreover, combining two rankings often outperforms the single ranking, showing that query expansion and neural re-ranking are highly complementary.\footnote{We have also experimented with combining BM25-QE, kNN and CE MAML + Query FT, however, have found it not to crucially outperform fusing only two rankings.} 
We analyze the intersection of the top documents between BM25-QE and the two re-rankers. We find that in more than $50\%$ of the queries in the test set, BM25-QE and the re-ranking model only agree on 5 or fewer documents in the top 20.\footnote{See Appendix \ref{sec:appendix-intersection} for details on the distribution.}

\subsection{Re-Ranking Ablations}\label{sec:ablation}
To gain further insights into where our performance improvements are coming from, we conduct a series of ablation studies, reported in Table~\ref{tab:ablations}. 

First, we ablate the influence of query expansion and the feedback documents on lexical retrieval. We retrieve only using the query and remove the feedback documents from the retrieval and evaluation, i.e. we use the residual collection, even though the feedback documents are not used. From the first section of Table~\ref{tab:ablations} we can observe a large performance drop. This shows that BM25-QE is successfully able to exploit the feedback documents and retrieve more relevant documents. 

For the kNN approach, we compare the performance by using only the query-document similarity for obtaining the relevance score (i.e. dropping the second term in Equation \ref{eq:knn}). On average this results in a drop of $5.6$ percentage points, proving the effectiveness of injecting feedback documents in the kNN re-ranking approach.

For the CE experiments, we ablate if optimizing only the bias layers compared to fully fine-tuning the model affects the performance. We, therefore, repeat our query fine-tuning experiment but optimize all parameters of the model. On average, optimizing only the biases results in a $0.8\%$ performance drop. However, the biases account only for 26k parameters, which is $0.11\%$ of the entire model. This result is in line with other research showing that optimizing only a small subset of parameters results in comparable performance \citep{pmlr-v97-houlsby19a,pfeiffer-etal-2020-adapterhub,ben-zaken-etal-2022-bitfit}. This finding supports the query fine-tuning applicability from a memory perspective. While there might be many queries in a deployed system, and therefore many fine-tuned models, the required memory would not grow significantly. Furthermore, the memory requirements could be further reduced by only fine-tuning biases of certain components \citep{ben-zaken-etal-2022-bitfit} or transformer layers \citep{ruckle-etal-2021-adapterdrop}.

Finally, we investigate the impact of meta-learning by comparing it with supervised training. We follow the same setup as in MAML but replace meta-learning with standard supervised learning. We find that MAML training results in $0.5\%$ improvement. We also note that supervised training is less stable than MAML. When increasing $k$, the performance intermittently drops (e.g. in TREC-Covid from $k=2 \rightarrow 4$ and TREC-News from $k=4 \rightarrow 8$), while MAML does not experience performance decreases.

\begin{table}[t!]
\small
\centering
\resizebox{\columnwidth}{!}{

\begin{tabularx}{\columnwidth}{@{}lYYYYY@{}}
\toprule
 & Robust & Covid & News & Touché & Avg. \\ \midrule
 
\multicolumn{6}{l}{\textit{BM25 without feedback documents}} \\ \midrule
& 0.0459 & 0.1615 & 0.0551 & 0.1052 & 0.0919 \\ \midrule
 
\multicolumn{6}{l}{\textit{kNN (Query Only)}} \\ \midrule
$k=2$ & 0.3531 & 0.6611  & 0.2537  & 0.1637  & 0.3579  \\  
$k=4$ & 0.3652 & 0.6486  & 0.2512  & 0.1649  & 0.3575  \\  
$k=8$ & 0.3677 & 0.6854  & 0.2578  & 0.1687  & 0.3699  \\ \cdashline{1-6} 
Avg. & 0.3620 & 0.6650  & 0.2542  & 0.1658  & 0.3618  \\ \midrule 

\multicolumn{6}{l}{\textit{CE Query-FT (full)}} \\ \midrule
$k=2$ & 0.4721 & 0.7168 & 0.3279 & 0.1797 & 0.4241 \\
$k=4$ & 0.5110 & 0.6872 & 0.3487 & 0.1858 & 0.4332 \\
$k=8$ & 0.5778 & 0.7644 & 0.3477 & 0.2021 & 0.4730 \\ \cdashline{1-6}
Avg.  & 0.5203 & 0.7228 & 0.3414 & 0.1892 & 0.4434 \\ \midrule

\multicolumn{6}{l}{\textit{CE supervised + Query-FT (bias)}} \\ \midrule
$k=2$ & 0.4540 & 0.7303 & 0.2716 & 0.2251 & 0.4203 \\
$k=4$ & 0.4896 & 0.7227 & 0.3657 & 0.2172 & 0.4488 \\
$k=8$ & 0.5353 & 0.7221 & 0.3390 & 0.2104 & 0.4517 \\ \cdashline{1-6}
Avg. & 0.4930 & 0.7250 & 0.3254 & 0.2176 & 0.4403 \\
\bottomrule
\end{tabularx}
}
\caption{nDCG@20 results of ablations studies on the test set. The first experiment shows BM25 without using query expansion and removing the feedback documents from the evaluation. The next experiment ablates
the performance of kNN by removing the influence of the relevant feedback documents. The third row shows results for fully fine-tuning the Cross-Encoder model, ablating fine-tuning only the bias layers. The last experiment ablates the MAML training, comparing it to standard supervised learning.}
\label{tab:ablations}
\end{table}

\subsection{Retrieval and Re-Ranking Latency}
\begin{figure}
    \centering
    \includegraphics[width=\columnwidth]{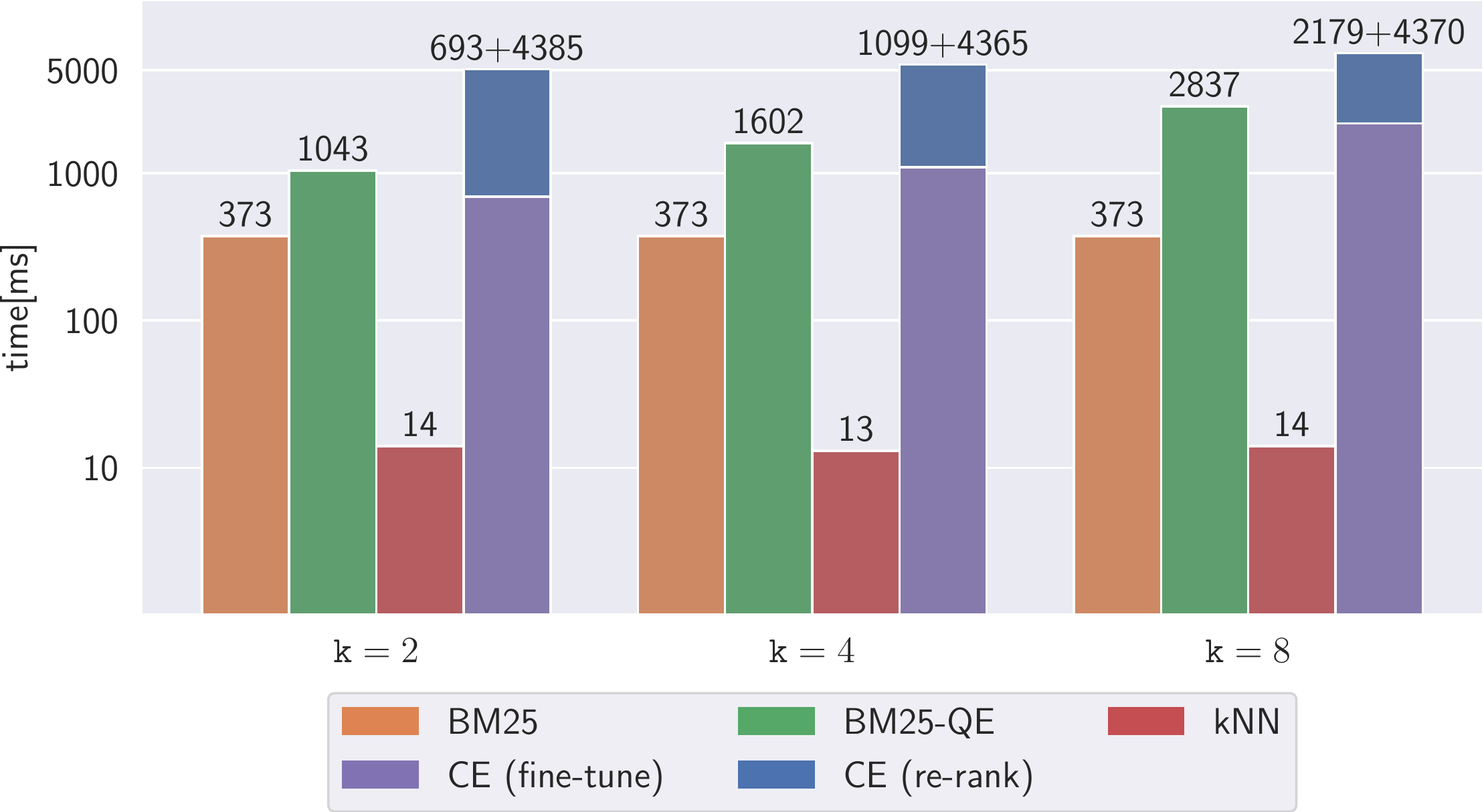}
    \caption{Average time in milliseconds (in log scale) for retrieval (BM25 and BM25-QE) and re-ranking (kNN and CE) 1000 documents. Average over all queries in the test sets. For the Cross-Encoder we separate the time for fine-tuning and re-ranking.}
    \label{fig:reranking-speed}
    \vspace{-3mm}
\end{figure}
Results for the speed performance are reported in Figure~\ref{fig:reranking-speed}. First, we note that performing query expansion does significantly increase retrieval speed. Depending on the number of feedback documents this is a $2.8$ ($k=2$) -- $7.6$ ($k=8$) fold increase over BM25 without query expansion.\footnote{For retrieval speed with varying number of expansion terms $e$ see Appendix \ref{sec:appendix-bm25-times}.}
\message{The column width is: \the\columnwidth}

For the re-ranking methods, we notice that the kNN approach is extremely fast. This is due to the fact that all heavy computations can be precomputed. This is promising since combining kNN and BM25-QE with rank fusion results in a $2.6\%$ performance improvement over BM25-QE alone, while not significantly adding any latency. In contrast, the Cross-Encoder model takes the longest time. However, the time for fine-tuning the model is only a fraction of the total time ($\approx 22\%$ on average). The retrieval latency can generally be traded-off with the ranking performance by retrieving and re-ranking fewer documents.

\section{Conclusion}
In this work, we introduced a few-shot learning task for incorporating relevance feedback in neural re-ranking models. We further transformed existing IR datasets into the few-shot setting. Most importantly, we have introduced different methods
for incorporating relevance feedback directly into neural re-ranking models. The proposed kNN approach is particularly computationally efficient, however, by itself, it cannot outperform BM25 with query expansion. Since the kNN method does not add significant latency to the re-ranking, it can be combined with BM25 query expansion, which outperforms the latter by $2.6\%$ nDCG@20. Our second re-ranking method based on a Cross-Encoder model performs on par with BM25 with query expansion. Regarding its latency, we show that fine-tuning on a query basis is feasible since a majority of the time is spent on re-ranking and not fine-tuning. Similar to kNN, performing rank fusion of the two approaches yields a high performance gain of $5.2\%$ nDCG@20. Advantageously, reciprocal rank fusion is very stable in our setting, mitigating weaknesses of individual model-task combinations.

\section{Limitations}
In this work, we investigate how relevance feedback can directly be incorporated into neural re-ranking models. While our best-performing approach improves the ranking performance by a large margin, it is inherently more computationally expensive compared with models that do not use any relevance feedback. We quantify this by reporting the latency of our approaches. The speed can be further reduced by re-ranking fewer documents, thereby trading off latency and performance. Further, we propose a kNN model that is computationally efficient and does not significantly add latency to query expansion models. Lastly, we recommend using our approach foremost in the information-seeking scenario, where users are particularly concerned about having accurate results rather than low latency.

For our methods, relevance feedback has to be explicitly collected from a user. While we believe in a information-seeking scenario users are more willing to provide explicit feedback, in this work, we did not explore using implicit or pseudo-relevance feedback. While this type of feedback is noisier, larger amounts are available.

In this work, we make use of simulated relevance feedback from existing relevance judgments. The re-ranked documents in the second stage will be biased toward the selected feedback documents. We leave to future work the integration of more diverse search results and investigation of position bias. However, we note that in preliminary experiments, we found that selecting random feedback documents from the first stage leads to worse performance. 

To keep the degrees of freedom in our experiments reasonable and to facilitate evaluation, we do not experiment with an iterative relevance feedback setting. We instead focus on a single round of relevance feedback but vary the number of feedback documents. While related work has shown that iterative relevance feedback can further improve retrieval, there are diminishing gains with every round \citep{bi2019iterative}.

Our best-performing approach requires a training dataset. Albeit small (depending on the task, the training dataset contains $22$ - $90$ queries), the model cannot be created without it. Since the model is targeted to a specific domain, we hypothesize that employing it on a different domain will result in worse performance than using the pre-trained model. To mitigate this, we also experiment with a model that does not use this intermediate fine-tuning step (CE Query-FT). Nevertheless, we encourage future work to look into combining unsupervised domain adaptation with our approaches to alleviate this limitation and potentially further improve performance. 

\section*{Acknowledgements}
We thank Max Glockner, Haritz Puerto, Rachneet Sachdeva, Gözde Gül Şahin, Thy Tran, and Kexin Wang for their insightful feedback on and reviews of earlier versions of the paper as well as the anonymous reviewers for their helpful comments and suggestions.

This work has been funded by the German Research Foundation (DFG) as part of the QASciInf project (grant GU 798/18-3) and the UKP-SQuARE project (grant GU 798/29-1). Furthermore, it has been supported by the German Federal Ministry of Education and Research and the Hessian Ministry of Higher Education, Research, Science and the Arts within their joint support of the National Research Center for Applied Cybersecurity ATHENE.

\bibliography{anthology,custom}

\begin{thebibliography}{54}
\expandafter\ifx\csname natexlab\endcsname\relax\def\natexlab#1{#1}\fi

\bibitem[{Ai et~al.(2018)Ai, Bi, Guo, and Croft}]{ai-2018-learning}
Qingyao Ai, Keping Bi, Jiafeng Guo, and W.~Bruce Croft. 2018.
\newblock \href {https://doi.org/10.1145/3209978.3209985} {Learning a deep
  listwise context model for ranking refinement}.
\newblock In \emph{The 41st International {ACM} {SIGIR} Conference on Research
  {\&} Development in Information Retrieval, {SIGIR} 2018, Ann Arbor, MI, USA,
  July 08-12, 2018}, pages 135--144. {ACM}.

\bibitem[{Ben~Zaken et~al.(2022)Ben~Zaken, Goldberg, and
  Ravfogel}]{ben-zaken-etal-2022-bitfit}
Elad Ben~Zaken, Yoav Goldberg, and Shauli Ravfogel. 2022.
\newblock \href {https://aclanthology.org/2022.acl-short.1} {{B}it{F}it: Simple
  parameter-efficient fine-tuning for transformer-based masked
  language-models}.
\newblock In \emph{Proceedings of the 60th Annual Meeting of the Association
  for Computational Linguistics (Volume 2: Short Papers)}, pages 1--9, Dublin,
  Ireland. Association for Computational Linguistics.

\bibitem[{Berger et~al.(2000)Berger, Caruana, Cohn, Freitag, and
  Mittal}]{10.1145/345508.345576}
Adam Berger, Rich Caruana, David Cohn, Dayne Freitag, and Vibhu Mittal. 2000.
\newblock \href {https://doi.org/10.1145/345508.345576} {Bridging the lexical
  chasm: Statistical approaches to answer-finding}.
\newblock In \emph{Proceedings of the 23rd Annual International ACM SIGIR
  Conference on Research and Development in Information Retrieval}, SIGIR '00,
  page 192–199, New York, NY, USA. Association for Computing Machinery.

\bibitem[{Bi et~al.(2019)Bi, Ai, and Croft}]{bi2019iterative}
Keping Bi, Qingyao Ai, and W.~Bruce Croft. 2019.
\newblock \href {https://doi.org/10.1007/978-3-030-15712-8\_36} {Iterative
  relevance feedback for answer passage retrieval with passage-level semantic
  match}.
\newblock In \emph{Advances in Information Retrieval - 41st European Conference
  on {IR} Research, {ECIR} 2019, Cologne, Germany, April 14-18, 2019,
  Proceedings, Part {I}}, volume 11437 of \emph{Lecture Notes in Computer
  Science}, pages 558--572. Springer.

\bibitem[{Bondarenko et~al.(2021)Bondarenko, Gienapp, Fr{\"{o}}be, Beloucif,
  Ajjour, Panchenko, Biemann, Stein, Wachsmuth, Potthast, and
  Hagen}]{bondarenko2020overview}
Alexander Bondarenko, Lukas Gienapp, Maik Fr{\"{o}}be, Meriem Beloucif, Yamen
  Ajjour, Alexander Panchenko, Chris Biemann, Benno Stein, Henning Wachsmuth,
  Martin Potthast, and Matthias Hagen. 2021.
\newblock \href {https://doi.org/10.1007/978-3-030-85251-1\_28} {Overview of
  touch{\'{e}} 2021: Argument retrieval}.
\newblock In \emph{Experimental {IR} Meets Multilinguality, Multimodality, and
  Interaction - 12th International Conference of the {CLEF} Association, {CLEF}
  2021, Virtual Event, September 21-24, 2021, Proceedings}, volume 12880 of
  \emph{Lecture Notes in Computer Science}, pages 450--467. Springer.

\bibitem[{Broder(2002)}]{broder_taxonomy}
Andrei~Z. Broder. 2002.
\newblock \href {https://doi.org/10.1145/792550.792552} {A taxonomy of web
  search}.
\newblock \emph{{SIGIR} Forum}, 36(2):3--10.

\bibitem[{Clarke et~al.(2008)Clarke, Kolla, Cormack, Vechtomova, Ashkan,
  B\"{u}ttcher, and MacKinnon}]{clarke-2008-novelty-diversity}
Charles~L.A. Clarke, Maheedhar Kolla, Gordon~V. Cormack, Olga Vechtomova, Azin
  Ashkan, Stefan B\"{u}ttcher, and Ian MacKinnon. 2008.
\newblock \href {https://doi.org/10.1145/1390334.1390446} {Novelty and
  diversity in information retrieval evaluation}.
\newblock In \emph{Proceedings of the 31st Annual International ACM SIGIR
  Conference on Research and Development in Information Retrieval}, SIGIR '08,
  page 659–666, New York, NY, USA. Association for Computing Machinery.

\bibitem[{Cormack et~al.(2009)Cormack, Clarke, and
  Buettcher}]{cromack-2009-reciprocal-rank-fusion}
Gordon~V. Cormack, Charles L~A Clarke, and Stefan Buettcher. 2009.
\newblock \href {https://doi.org/10.1145/1571941.1572114} {Reciprocal rank
  fusion outperforms condorcet and individual rank learning methods}.
\newblock In \emph{Proceedings of the 32nd International ACM SIGIR Conference
  on Research and Development in Information Retrieval}, SIGIR '09, page
  758–759, New York, NY, USA. Association for Computing Machinery.

\bibitem[{Dasigi et~al.(2021)Dasigi, Lo, Beltagy, Cohan, Smith, and
  Gardner}]{dasigi-etal-2021-dataset}
Pradeep Dasigi, Kyle Lo, Iz~Beltagy, Arman Cohan, Noah~A. Smith, and Matt
  Gardner. 2021.
\newblock \href {https://doi.org/10.18653/v1/2021.naacl-main.365} {A dataset of
  information-seeking questions and answers anchored in research papers}.
\newblock In \emph{Proceedings of the 2021 Conference of the North American
  Chapter of the Association for Computational Linguistics: Human Language
  Technologies}, pages 4599--4610, Online. Association for Computational
  Linguistics.

\bibitem[{Devlin et~al.(2019)Devlin, Chang, Lee, and
  Toutanova}]{devlin-etal-2019-bert}
Jacob Devlin, Ming-Wei Chang, Kenton Lee, and Kristina Toutanova. 2019.
\newblock \href {https://doi.org/10.18653/v1/N19-1423} {{BERT}: Pre-training of
  deep bidirectional transformers for language understanding}.
\newblock In \emph{Proceedings of the 2019 Conference of the North {A}merican
  Chapter of the Association for Computational Linguistics: Human Language
  Technologies, Volume 1 (Long and Short Papers)}, pages 4171--4186,
  Minneapolis, Minnesota. Association for Computational Linguistics.

\bibitem[{Finn et~al.(2017)Finn, Abbeel, and Levine}]{pmlr-v70-finn17a}
Chelsea Finn, Pieter Abbeel, and Sergey Levine. 2017.
\newblock \href {https://proceedings.mlr.press/v70/finn17a.html}
  {Model-agnostic meta-learning for fast adaptation of deep networks}.
\newblock In \emph{Proceedings of the 34th International Conference on Machine
  Learning}, volume~70 of \emph{Proceedings of Machine Learning Research},
  pages 1126--1135. PMLR.

\bibitem[{Furnas et~al.(1987)Furnas, Landauer, Gomez, and
  Dumais}]{furnas-1987-vocab}
George~W. Furnas, Thomas~K. Landauer, Louis~M. Gomez, and Susan~T. Dumais.
  1987.
\newblock \href {https://doi.org/10.1145/32206.32212} {The vocabulary problem
  in human-system communication}.
\newblock \emph{Commun. {ACM}}, 30(11):964--971.

\bibitem[{Gillick et~al.(2019)Gillick, Kulkarni, Lansing, Presta, Baldridge,
  Ie, and Garcia-Olano}]{gillick-etal-2019-learning}
Daniel Gillick, Sayali Kulkarni, Larry Lansing, Alessandro Presta, Jason
  Baldridge, Eugene Ie, and Diego Garcia-Olano. 2019.
\newblock \href {https://doi.org/10.18653/v1/K19-1049} {Learning dense
  representations for entity retrieval}.
\newblock In \emph{Proceedings of the 23rd Conference on Computational Natural
  Language Learning (CoNLL)}, pages 528--537, Hong Kong, China. Association for
  Computational Linguistics.

\bibitem[{Guu et~al.(2020)Guu, Lee, Tung, Pasupat, and Chang}]{realm}
Kelvin Guu, Kenton Lee, Zora Tung, Panupong Pasupat, and Ming{-}Wei Chang.
  2020.
\newblock \href {http://arxiv.org/abs/2002.08909} {{REALM:} retrieval-augmented
  language model pre-training}.
\newblock \emph{CoRR}, abs/2002.08909.

\bibitem[{Gysel and de~Rijke(2018)}]{VanGysel2018pytreceval}
Christophe~Van Gysel and Maarten de~Rijke. 2018.
\newblock \href {https://doi.org/10.1145/3209978.3210065} {Pytrec{\_}eval: An
  extremely fast python interface to trec{\_}eval}.
\newblock In \emph{The 41st International {ACM} {SIGIR} Conference on Research
  {\&} Development in Information Retrieval, {SIGIR} 2018, Ann Arbor, MI, USA,
  July 08-12, 2018}, pages 873--876. {ACM}.

\bibitem[{Hofst{\"{a}}tter et~al.(2020)Hofst{\"{a}}tter, Althammer,
  Schr{\"{o}}der, Sertkan, and Hanbury}]{hofstatter2020improving}
Sebastian Hofst{\"{a}}tter, Sophia Althammer, Michael Schr{\"{o}}der, Mete
  Sertkan, and Allan Hanbury. 2020.
\newblock \href {http://arxiv.org/abs/2010.02666} {Improving efficient neural
  ranking models with cross-architecture knowledge distillation}.
\newblock \emph{CoRR}, abs/2010.02666.

\bibitem[{Hofst\"{a}tter et~al.(2021)Hofst\"{a}tter, Lin, Yang, Lin, and
  Hanbury}]{hoffstaetter-2021-effiecently}
Sebastian Hofst\"{a}tter, Sheng-Chieh Lin, Jheng-Hong Yang, Jimmy Lin, and
  Allan Hanbury. 2021.
\newblock \href {https://doi.org/10.1145/3404835.3462891} {Efficiently teaching
  an effective dense retriever with balanced topic aware sampling}.
\newblock In \emph{Proceedings of the 44th International ACM SIGIR Conference
  on Research and Development in Information Retrieval}, SIGIR '21, page
  113–122, New York, NY, USA. Association for Computing Machinery.

\bibitem[{Houlsby et~al.(2019)Houlsby, Giurgiu, Jastrzebski, Morrone,
  De~Laroussilhe, Gesmundo, Attariyan, and Gelly}]{pmlr-v97-houlsby19a}
Neil Houlsby, Andrei Giurgiu, Stanislaw Jastrzebski, Bruna Morrone, Quentin
  De~Laroussilhe, Andrea Gesmundo, Mona Attariyan, and Sylvain Gelly. 2019.
\newblock \href {https://proceedings.mlr.press/v97/houlsby19a.html}
  {Parameter-efficient transfer learning for {NLP}}.
\newblock In \emph{Proceedings of the 36th International Conference on Machine
  Learning}, volume~97 of \emph{Proceedings of Machine Learning Research},
  pages 2790--2799. PMLR.

\bibitem[{J{\"{a}}rvelin and Kek{\"{a}}l{\"{a}}inen(2000)}]{jarvelin2000ir}
Kalervo J{\"{a}}rvelin and Jaana Kek{\"{a}}l{\"{a}}inen. 2000.
\newblock \href {https://doi.org/10.1145/345508.345545} {{IR} evaluation
  methods for retrieving highly relevant documents}.
\newblock In \emph{{SIGIR} 2000: Proceedings of the 23rd Annual International
  {ACM} {SIGIR} Conference on Research and Development in Information
  Retrieval, July 24-28, 2000, Athens, Greece}, pages 41--48. {ACM}.

\bibitem[{Joachims(2002)}]{joachims-2002-clickthrough}
Thorsten Joachims. 2002.
\newblock \href {https://doi.org/10.1145/775047.775067} {Optimizing search
  engines using clickthrough data}.
\newblock In \emph{Proceedings of the Eighth ACM SIGKDD International
  Conference on Knowledge Discovery and Data Mining}, KDD '02, page 133–142,
  New York, NY, USA. Association for Computing Machinery.

\bibitem[{Johnson et~al.(2021)Johnson, Douze, and
  J{\'{e}}gou}]{johnson2019billion}
Jeff Johnson, Matthijs Douze, and Herv{\'{e}} J{\'{e}}gou. 2021.
\newblock \href {https://doi.org/10.1109/TBDATA.2019.2921572} {Billion-scale
  similarity search with gpus}.
\newblock \emph{{IEEE} Trans. Big Data}, 7(3):535--547.

\bibitem[{Karpukhin et~al.(2020)Karpukhin, Oguz, Min, Lewis, Wu, Edunov, Chen,
  and Yih}]{karpukhin-etal-2020-dense}
Vladimir Karpukhin, Barlas Oguz, Sewon Min, Patrick Lewis, Ledell Wu, Sergey
  Edunov, Danqi Chen, and Wen-tau Yih. 2020.
\newblock \href {https://doi.org/10.18653/v1/2020.emnlp-main.550} {Dense
  passage retrieval for open-domain question answering}.
\newblock In \emph{Proceedings of the 2020 Conference on Empirical Methods in
  Natural Language Processing (EMNLP)}, pages 6769--6781, Online. Association
  for Computational Linguistics.

\bibitem[{Khattab and Zaharia(2020)}]{khattab-zaharia-2020-colbert}
Omar Khattab and Matei Zaharia. 2020.
\newblock \href {https://doi.org/10.1145/3397271.3401075} {\emph{ColBERT:
  Efficient and Effective Passage Search via Contextualized Late Interaction
  over BERT}}, page 39–48. Association for Computing Machinery, New York, NY,
  USA.

\bibitem[{Kwiatkowski et~al.(2019)Kwiatkowski, Palomaki, Redfield, Collins,
  Parikh, Alberti, Epstein, Polosukhin, Devlin, Lee, Toutanova, Jones, Kelcey,
  Chang, Dai, Uszkoreit, Le, and Petrov}]{kwiatkowski-etal-2019-natural}
Tom Kwiatkowski, Jennimaria Palomaki, Olivia Redfield, Michael Collins, Ankur
  Parikh, Chris Alberti, Danielle Epstein, Illia Polosukhin, Jacob Devlin,
  Kenton Lee, Kristina Toutanova, Llion Jones, Matthew Kelcey, Ming-Wei Chang,
  Andrew~M. Dai, Jakob Uszkoreit, Quoc Le, and Slav Petrov. 2019.
\newblock \href {https://doi.org/10.1162/tacl_a_00276} {Natural questions: A
  benchmark for question answering research}.
\newblock \emph{Transactions of the Association for Computational Linguistics},
  7:452--466.

\bibitem[{Lavrenko and Croft(2001)}]{10.1145/383952.383972}
Victor Lavrenko and W.~Bruce Croft. 2001.
\newblock \href {https://doi.org/10.1145/383952.383972} {Relevance based
  language models}.
\newblock In \emph{Proceedings of the 24th Annual International ACM SIGIR
  Conference on Research and Development in Information Retrieval}, SIGIR '01,
  page 120–127, New York, NY, USA. Association for Computing Machinery.

\bibitem[{Li et~al.(2018)Li, Sun, He, Wang, Hui, Yates, Sun, and
  Xu}]{li-etal-2018-nprf}
Canjia Li, Yingfei Sun, Ben He, Le~Wang, Kai Hui, Andrew Yates, Le~Sun, and
  Jungang Xu. 2018.
\newblock \href {https://doi.org/10.18653/v1/D18-1478} {{NPRF}: A neural pseudo
  relevance feedback framework for ad-hoc information retrieval}.
\newblock In \emph{Proceedings of the 2018 Conference on Empirical Methods in
  Natural Language Processing}, pages 4482--4491, Brussels, Belgium.
  Association for Computational Linguistics.

\bibitem[{Lin(2019)}]{lin2019simplest}
Jimmy Lin. 2019.
\newblock \href {http://arxiv.org/abs/1904.08861} {The simplest thing that can
  possibly work: Pseudo-relevance feedback using text classification}.
\newblock \emph{CoRR}, abs/1904.08861.

\bibitem[{Locke and Zuccon(2018)}]{locke-zuccon-test-2018}
Daniel Locke and Guido Zuccon. 2018.
\newblock \href {https://doi.org/10.1145/3209978.3210161} {A test collection
  for evaluating legal case law search}.
\newblock In \emph{The 41st International ACM SIGIR Conference on Research \&
  Development in Information Retrieval}, SIGIR '18, page 1261–1264, New York,
  NY, USA. Association for Computing Machinery.

\bibitem[{Loshchilov and Hutter(2019)}]{adamW}
Ilya Loshchilov and Frank Hutter. 2019.
\newblock \href {https://openreview.net/forum?id=Bkg6RiCqY7} {Decoupled weight
  decay regularization}.
\newblock In \emph{7th International Conference on Learning Representations,
  {ICLR} 2019, New Orleans, LA, USA, May 6-9, 2019}. OpenReview.net.

\bibitem[{Naseri et~al.(2021)Naseri, Dalton, Yates, and
  Allan}]{shahrzad-etal-2021-ceqe}
Shahrzad Naseri, Jeffrey Dalton, Andrew Yates, and James Allan. 2021.
\newblock \href {https://doi.org/10.1007/978-3-030-72113-8_31} {Ceqe:
  Contextualized embeddings for query expansion}.
\newblock In \emph{Advances in Information Retrieval: 43rd European Conference
  on IR Research, ECIR 2021, Virtual Event, March 28 – April 1, 2021,
  Proceedings, Part I}, page 467–482, Berlin, Heidelberg. Springer-Verlag.

\bibitem[{Nguyen et~al.(2016)Nguyen, Rosenberg, Song, Gao, Tiwary, Majumder,
  and Deng}]{nguyen2016ms}
Tri Nguyen, Mir Rosenberg, Xia Song, Jianfeng Gao, Saurabh Tiwary, Rangan
  Majumder, and Li~Deng. 2016.
\newblock \href {http://ceur-ws.org/Vol-1773/CoCoNIPS\_2016\_paper9.pdf} {{MS}
  {MARCO:} {A} human generated machine reading comprehension dataset}.
\newblock 1773.

\bibitem[{Ni et~al.(2021)Ni, Qu, Lu, Dai, {\'{A}}brego, Ma, Zhao, Luan, Hall,
  Chang, and Yang}]{ni2021large}
Jianmo Ni, Chen Qu, Jing Lu, Zhuyun Dai, Gustavo~Hern{\'{a}}ndez {\'{A}}brego,
  Ji~Ma, Vincent~Y. Zhao, Yi~Luan, Keith~B. Hall, Ming{-}Wei Chang, and Yinfei
  Yang. 2021.
\newblock \href {http://arxiv.org/abs/2112.07899} {Large dual encoders are
  generalizable retrievers}.
\newblock \emph{CoRR}, abs/2112.07899.

\bibitem[{Nogueira and Cho(2019)}]{nogueira-cho-2019-reranking}
Rodrigo Nogueira and Kyunghyun Cho. 2019.
\newblock \href {http://arxiv.org/abs/1901.04085} {Passage re-ranking with
  {BERT}}.
\newblock \emph{CoRR}, abs/1901.04085.

\bibitem[{Pfeiffer et~al.(2020)Pfeiffer, R{\"u}ckl{\'e}, Poth, Kamath,
  Vuli{\'c}, Ruder, Cho, and Gurevych}]{pfeiffer-etal-2020-adapterhub}
Jonas Pfeiffer, Andreas R{\"u}ckl{\'e}, Clifton Poth, Aishwarya Kamath, Ivan
  Vuli{\'c}, Sebastian Ruder, Kyunghyun Cho, and Iryna Gurevych. 2020.
\newblock \href {https://doi.org/10.18653/v1/2020.emnlp-demos.7}
  {{A}dapter{H}ub: A framework for adapting transformers}.
\newblock In \emph{Proceedings of the 2020 Conference on Empirical Methods in
  Natural Language Processing: System Demonstrations}, pages 46--54, Online.
  Association for Computational Linguistics.

\bibitem[{Reimers and Gurevych(2019)}]{reimers-gurevych-2019-sentence}
Nils Reimers and Iryna Gurevych. 2019.
\newblock \href {https://doi.org/10.18653/v1/D19-1410} {Sentence-{BERT}:
  Sentence embeddings using {S}iamese {BERT}-networks}.
\newblock In \emph{Proceedings of the 2019 Conference on Empirical Methods in
  Natural Language Processing and the 9th International Joint Conference on
  Natural Language Processing (EMNLP-IJCNLP)}, pages 3982--3992, Hong Kong,
  China. Association for Computational Linguistics.

\bibitem[{Robertson and Zaragoza(2009)}]{robertson-zaragoza-2009-bm25}
Stephen Robertson and Hugo Zaragoza. 2009.
\newblock \href {https://doi.org/10.1561/1500000019} {The probabilistic
  relevance framework: Bm25 and beyond}.
\newblock \emph{Found. Trends Inf. Retr.}, 3(4):333–389.

\bibitem[{Rocchio(1971)}]{rocchio1971relevance}
Joseph Rocchio. 1971.
\newblock Relevance feedback in information retrieval.
\newblock \emph{The Smart retrieval system-experiments in automatic document
  processing}, pages 313--323.

\bibitem[{R{\"u}ckl{\'e} et~al.(2021)R{\"u}ckl{\'e}, Geigle, Glockner, Beck,
  Pfeiffer, Reimers, and Gurevych}]{ruckle-etal-2021-adapterdrop}
Andreas R{\"u}ckl{\'e}, Gregor Geigle, Max Glockner, Tilman Beck, Jonas
  Pfeiffer, Nils Reimers, and Iryna Gurevych. 2021.
\newblock \href {https://doi.org/10.18653/v1/2021.emnlp-main.626}
  {{AdapterDrop}: {O}n the efficiency of adapters in transformers}.
\newblock In \emph{Proceedings of the 2021 Conference on Empirical Methods in
  Natural Language Processing}, pages 7930--7946, Online and Punta Cana,
  Dominican Republic. Association for Computational Linguistics.

\bibitem[{Salton and Buckley(1990)}]{salton1990improving}
Gerard Salton and Chris Buckley. 1990.
\newblock Improving retrieval performance by relevance feedback.
\newblock \emph{Journal of the American society for information science},
  41(4):288--297.

\bibitem[{Schurman and Brutlag(2009)}]{schurman2009performance}
Eric Schurman and Jake Brutlag. 2009.
\newblock Performance related changes and their user impact.
\newblock Presented at Velocity Web Performance and Operations Conference.

\bibitem[{Snell et~al.(2017)Snell, Swersky, and Zemel}]{NIPS2017_cb8da676}
Jake Snell, Kevin Swersky, and Richard~S. Zemel. 2017.
\newblock \href
  {https://proceedings.neurips.cc/paper/2017/hash/cb8da6767461f2812ae4290eac7cbc42-Abstract.html}
  {Prototypical networks for few-shot learning}.
\newblock In \emph{Advances in Neural Information Processing Systems 30: Annual
  Conference on Neural Information Processing Systems 2017, December 4-9, 2017,
  Long Beach, CA, {USA}}, pages 4077--4087.

\bibitem[{Soboroff et~al.(2018)Soboroff, Huang, and Harman}]{soboroff2018trec}
Ian Soboroff, Shudong Huang, and Donna Harman. 2018.
\newblock \href {https://trec.nist.gov/pubs/trec27/papers/Overview-News.pdf}
  {{TREC} 2018 news track overview}.
\newblock In \emph{Proceedings of the Twenty-Seventh Text REtrieval Conference,
  {TREC} 2018, Gaithersburg, Maryland, USA, November 14-16, 2018}, volume
  500-331 of \emph{{NIST} Special Publication}. National Institute of Standards
  and Technology {(NIST)}.

\bibitem[{Thakur et~al.(2021)Thakur, Reimers, R\"{u}ckl\'{e}, Srivastava, and
  Gurevych}]{thakur2021beir}
Nandan Thakur, Nils Reimers, Andreas R\"{u}ckl\'{e}, Abhishek Srivastava, and
  Iryna Gurevych. 2021.
\newblock \href
  {https://datasets-benchmarks-proceedings.neurips.cc/paper/2021/file/65b9eea6e1cc6bb9f0cd2a47751a186f-Paper-round2.pdf}
  {Beir: A heterogeneous benchmark for zero-shot evaluation of information
  retrieval models}.
\newblock In \emph{Proceedings of the Neural Information Processing Systems
  Track on Datasets and Benchmarks}, volume~1.

\bibitem[{Voorhees et~al.(2021)Voorhees, Alam, Bedrick, Demner-Fushman, Hersh,
  Lo, Roberts, Soboroff, and Wang}]{voorhees2021trec}
Ellen Voorhees, Tasmeer Alam, Steven Bedrick, Dina Demner-Fushman, William~R.
  Hersh, Kyle Lo, Kirk Roberts, Ian Soboroff, and Lucy~Lu Wang. 2021.
\newblock \href {https://doi.org/10.1145/3451964.3451965} {Trec-covid:
  Constructing a pandemic information retrieval test collection}.
\newblock \emph{SIGIR Forum}, 54(1).

\bibitem[{Voorhees(2004)}]{voorhees2004overview}
Ellen~M. Voorhees. 2004.
\newblock \href {http://trec.nist.gov/pubs/trec13/papers/ROBUST.OVERVIEW.pdf}
  {Overview of the {TREC} 2004 robust track}.
\newblock In \emph{Proceedings of the Thirteenth Text REtrieval Conference,
  {TREC} 2004, Gaithersburg, Maryland, USA, November 16-19, 2004}, volume
  500-261 of \emph{{NIST} Special Publication}. National Institute of Standards
  and Technology {(NIST)}.

\bibitem[{Wang et~al.(2022)Wang, Thakur, Reimers, and
  Gurevych}]{wang-etal-2022-gpl}
Kexin Wang, Nandan Thakur, Nils Reimers, and Iryna Gurevych. 2022.
\newblock \href {https://doi.org/10.18653/v1/2022.naacl-main.168} {{GPL}:
  Generative pseudo labeling for unsupervised domain adaptation of dense
  retrieval}.
\newblock In \emph{Proceedings of the 2022 Conference of the North American
  Chapter of the Association for Computational Linguistics: Human Language
  Technologies}, pages 2345--2360, Seattle, United States. Association for
  Computational Linguistics.

\bibitem[{Wang et~al.(2020{\natexlab{a}})Wang, Lo, Chandrasekhar, Reas, Yang,
  Burdick, Eide, Funk, Katsis, Kinney, Li, Liu, Merrill, Mooney, Murdick,
  Rishi, Sheehan, Shen, Stilson, Wade, Wang, Wang, Wilhelm, Xie, Raymond, Weld,
  Etzioni, and Kohlmeier}]{wang-etal-2020-cord}
Lucy~Lu Wang, Kyle Lo, Yoganand Chandrasekhar, Russell Reas, Jiangjiang Yang,
  Doug Burdick, Darrin Eide, Kathryn Funk, Yannis Katsis, Rodney~Michael
  Kinney, Yunyao Li, Ziyang Liu, William Merrill, Paul Mooney, Dewey~A.
  Murdick, Devvret Rishi, Jerry Sheehan, Zhihong Shen, Brandon Stilson, Alex~D.
  Wade, Kuansan Wang, Nancy Xin~Ru Wang, Christopher Wilhelm, Boya Xie,
  Douglas~M. Raymond, Daniel~S. Weld, Oren Etzioni, and Sebastian Kohlmeier.
  2020{\natexlab{a}}.
\newblock \href {https://aclanthology.org/2020.nlpcovid19-acl.1} {{CORD-19}:
  The {COVID-19} open research dataset}.
\newblock In \emph{Proceedings of the 1st Workshop on {NLP} for {COVID-19} at
  {ACL} 2020}, Online. Association for Computational Linguistics.

\bibitem[{Wang et~al.(2020{\natexlab{b}})Wang, Wei, Dong, Bao, Yang, and
  Zhou}]{NEURIPS2020_3f5ee243}
Wenhui Wang, Furu Wei, Li~Dong, Hangbo Bao, Nan Yang, and Ming Zhou.
  2020{\natexlab{b}}.
\newblock \href
  {https://proceedings.neurips.cc/paper/2020/hash/3f5ee243547dee91fbd053c1c4a845aa-Abstract.html}
  {Minilm: Deep self-attention distillation for task-agnostic compression of
  pre-trained transformers}.
\newblock In \emph{Advances in Neural Information Processing Systems 33: Annual
  Conference on Neural Information Processing Systems 2020, NeurIPS 2020,
  December 6-12, 2020, virtual}.

\bibitem[{Wu and Fang(2013)}]{wu2013incremental}
Hao Wu and Hui Fang. 2013.
\newblock \href {https://doi.org/10.1145/2484028.2484051} {An incremental
  approach to efficient pseudo-relevance feedback}.
\newblock In \emph{The 36th International {ACM} {SIGIR} conference on research
  and development in Information Retrieval, {SIGIR} '13, Dublin, Ireland - July
  28 - August 01, 2013}, pages 553--562. {ACM}.

\bibitem[{Xiong et~al.(2021)Xiong, Xiong, Li, Tang, Liu, Bennett, Ahmed, and
  Overwijk}]{xiong2021approximate}
Lee Xiong, Chenyan Xiong, Ye~Li, Kwok-Fung Tang, Jialin Liu, Paul~N. Bennett,
  Junaid Ahmed, and Arnold Overwijk. 2021.
\newblock \href {https://openreview.net/forum?id=zeFrfgyZln} {Approximate
  nearest neighbor negative contrastive learning for dense text retrieval}.
\newblock In \emph{International Conference on Learning Representations}.

\bibitem[{Yu et~al.(2021)Yu, Xiong, and Callan}]{yu-2021-improving}
HongChien Yu, Chenyan Xiong, and Jamie Callan. 2021.
\newblock \href {https://doi.org/10.1145/3459637.3482124} {Improving query
  representations for dense retrieval with pseudo relevance feedback}.
\newblock In \emph{Proceedings of the 30th ACM International Conference on
  Information \& Knowledge Management}, CIKM '21, page 3592–3596, New York,
  NY, USA. Association for Computing Machinery.

\bibitem[{Zhai and Lafferty(2001)}]{zhai2001model}
ChengXiang Zhai and John~D. Lafferty. 2001.
\newblock \href {https://doi.org/10.1145/502585.502654} {Model-based feedback
  in the language modeling approach to information retrieval}.
\newblock In \emph{Proceedings of the 2001 {ACM} {CIKM} International
  Conference on Information and Knowledge Management, Atlanta, Georgia, USA,
  November 5-10, 2001}, pages 403--410. {ACM}.

\bibitem[{Zhang et~al.(2021)Zhang, Ma, Shi, and Lin}]{zhang-etal-2021-mr}
Xinyu Zhang, Xueguang Ma, Peng Shi, and Jimmy Lin. 2021.
\newblock \href {https://doi.org/10.18653/v1/2021.mrl-1.12} {Mr. {T}y{D}i: A
  multi-lingual benchmark for dense retrieval}.
\newblock In \emph{Proceedings of the 1st Workshop on Multilingual
  Representation Learning}, pages 127--137, Punta Cana, Dominican Republic.
  Association for Computational Linguistics.

\bibitem[{Zheng et~al.(2020)Zheng, Hui, He, Han, Sun, and
  Yates}]{zheng-etal-2020-bert}
Zhi Zheng, Kai Hui, Ben He, Xianpei Han, Le~Sun, and Andrew Yates. 2020.
\newblock \href {https://doi.org/10.18653/v1/2020.findings-emnlp.424}
  {{BERT-QE}: {C}ontextualized {Q}uery {E}xpansion for {D}ocument
  {R}e-ranking}.
\newblock In \emph{Findings of the Association for Computational Linguistics:
  EMNLP 2020}, pages 4718--4728, Online. Association for Computational
  Linguistics.

\end{thebibliography}
\bibliographystyle{acl_natbib}

\clearpage

\appendix
\section{Dataset Details}\label{sec:appendix-datasets}
\paragraph{Robust04 \citep{voorhees2004overview}} is a dataset initially created to investigate the performance of poorly performing queries. Thereby, a collection with many judgments per query has been created and has since been used to test the robustness of IR models. We use the description field of the queries, which is a question or a single sentence of the search intent. The document collection contains news articles.

\paragraph{TREC-Covid \citep{voorhees2021trec}} is an IR dataset in the biomedical domain consisting of questions about Coronavirus and scientific articles as document collection. It was collected in five iterative rounds. We use the question from the query set along with the documents from the COVID-19 Open Research Dataset \citep{wang-etal-2020-cord}.\footnote{We use the snapshot from 16-JULY-2020} Documents are constructed by concatenating the title and abstract. Further, we remove exact duplicates from the feedback documents. In TREC-Covid, the documents are judged as relevant, partially relevant, or non-relevant. For the feedback documents, we consider only the relevant and non-relevant ones but include also partially relevant ones for evaluation.

\paragraph{TREC-News \citep{soboroff2018trec}} is an Information Retrieval task based on a corpus provided by the Washington Post. We use the 2019 background linking task. In this setup, the goal is to find other relevant news articles that provide background information or further reading on a subject and help the user contextualize the current article. To have a concise query, we use the titles as query.

\paragraph{Webis-Touché 2020 \citep{bondarenko2020overview}} is an argument retrieval dataset based on the args.me\footnote{\href{www.args.me/api-en.html}{args.me/api-en.html}} corpus containing arguments scraped from debate websites.\footnote{\href{www.debate.org}{debate.org}, \href{www.debatepedia.org}{debatepedia.org}, \href{www.debatewise.org}{debatewise.org}, \href{www.idebate.org}{idebate.org}} Queries are formulated as questions. The dataset contains fine-grained annotations of documents on a scale from 0-7. We select documents with a relevance of at least 3 for our relevant feedback documents. Since Webis-Touché only contains very few non-relevant documents (i.e. relevance of 0), we augment them using BM25 negatives \citep{karpukhin-etal-2020-dense}, by selecting non-judged documents after rank 100 for the non-relevant feedback documents.

\section{1\textsuperscript{st} Stage Retrieval}\label{sec:first-stage-retrieval-results}

Table~\ref{tab:bm25-results} shows BM25 retrieval performance with the relevance feedback documents removed ($F=\text{\xmark}$) or included ($F=\text{\cmark}$) in the retrieval results and evaluation.

\begin{table}[h!]
\centering
\resizebox{\columnwidth}{!}{
\begin{tabular}{@{}llccc@{}}
\toprule
 & $F$ & nDCG@20 & R@100 & R@1000\\ \midrule
\multirow{2}{*}{Robust} 
 & \cmark & 0.3292 & 0.2256 & 0.5021 \\
 & \xmark & 0.0459 & 0.1488 & 0.4464 \\ \midrule
\multirow{2}{*}{Covid} 
 & \cmark & 0.5597 & 0.0963 & 0.3686 \\
 & \xmark & 0.1615 & 0.0783 & 0.3556 \\ \midrule
\multirow{2}{*}{News} 
 & \cmark & 0.2993 & 0.4039 & 0.7406  \\
 & \xmark & 0.0551 & 0.3137 & 0.7017 \\ \midrule
\multirow{2}{*}{Touché} 
 & \cmark & 0.5134 & 0.4067 & 0.6461 \\
 & \xmark & 0.1052 & 0.2887 & 0.5717  \\ \midrule
\multirow{2}{*}{Avg.} 
 & \cmark & 0.4254 & 0.2831 & 0.5644\\
 & \xmark & 0.0919 & 0.2074 & 0.5188 \\ \bottomrule
\end{tabular}
}
\caption{Retrieval results using BM25 on the test set with the query only. For $F$ = \xmark \ the feedback documents have been removed from the retrieved documents and the ground truth for computing the evaluation metrics.}
\label{tab:bm25-results}
\end{table}

\clearpage 
\onecolumn

\section{Zero-Shot Baselines}\label{sec:zero-shot-baselines}
\begin{table}[h!]
\small
\begin{minipage}{.475\linewidth}
\centering
\subfloat[nDCG@20 on the validation set.]{
\begin{tabularx}{\columnwidth}{@{}lYYYYY@{}}
\toprule
 & Robust & Covid & News & Touche & Avg. \\ \midrule
 
 \multicolumn{6}{l}{\textit{DPR-single} (109M) \citep{karpukhin-etal-2020-dense}} \\ \midrule
$k=2$ & 0.1335 & 0.4232  & 0.1913  & 0.0510  & 0.1997  \\  
$k=4$ & 0.1254 & 0.4359  & 0.1922  & 0.0547  & 0.2020  \\  
$k=8$ & 0.1510 & 0.4749  & 0.1981  & 0.0603  & 0.2211  \\ \cdashline{1-6} 
Avg. & 0.1366 & 0.4447  & 0.1938  & 0.0553  & 0.2076  \\ \midrule 

\multicolumn{6}{l}{\textit{DPR-multi} (109M) \citep{karpukhin-etal-2020-dense}} \\ \midrule
$k=2$ & 0.2570 & 0.4466  & 0.2191  & 0.0917  & 0.2536  \\  
$k=4$ & 0.2662 & 0.4312  & 0.2156  & 0.0924  & 0.2513  \\  
$k=8$ & 0.2690 & 0.4431  & 0.2188  & 0.0959  & 0.2567  \\ \cdashline{1-6} 
Avg. & 0.2641 & 0.4403  & 0.2178  & 0.0933  & 0.2539  \\ \midrule

\multicolumn{6}{l}{\textit{ANCE} (125M) \citep{xiong2021approximate}} \\ \midrule
$k=2$ & 0.3481 & 0.6671  & 0.3061  & 0.1511  & 0.3681  \\  
$k=4$ & 0.3574 & 0.6825  & 0.3064  & 0.1497  & 0.3740  \\  
$k=8$ & 0.3662 & 0.6834  & 0.3137  & 0.1544  & 0.3794  \\ \cdashline{1-6} 
Avg. & 0.3572 & 0.6777  & 0.3087  & 0.1517  & 0.3738  \\ \midrule

\multicolumn{6}{l}{\textit{MiniLM} (23M) \citep{NEURIPS2020_3f5ee243}} \\ \midrule
$k=2$ & 0.3824 & 0.5795  & 0.2800  & 0.1433  & 0.3463  \\  
$k=4$ & 0.3843 & 0.5847  & 0.2831  & 0.1570  & 0.3523  \\  
$k=8$ & 0.4044 & 0.6265  & 0.2856  & 0.1668  & 0.3708  \\ \cdashline{1-6} 
Avg. & 0.3903 & 0.5969  & 0.2829  & 0.1557  & 0.3565  \\ \midrule 

\multicolumn{6}{l}{\textit{TAS-B} (66M) \citep{hoffstaetter-2021-effiecently}} \\ \midrule
$k=2$ & 0.3637 & 0.6475  & 0.2549  & 0.1397  & 0.3515  \\  
$k=4$ & 0.3666 & 0.6568  & 0.2612  & 0.1570  & 0.3604  \\  
$k=8$ & 0.3736 & 0.6607  & 0.2576  & 0.1553  & 0.3618  \\ \cdashline{1-6} 
Avg. & 0.3680 & 0.6550  & 0.2579  & 0.1507  & 0.3579  \\ \midrule 

\multicolumn{6}{l}{\textit{GPL} (66M) \citep{wang-etal-2022-gpl}} \\ \midrule
$k=2$ & 0.3741 & 0.6626  & 0.3049  & 0.1588  & 0.3751  \\  
$k=4$ & 0.3810 & 0.6781  & 0.3090  & 0.1646  & 0.3832  \\  
$k=8$ & 0.3880 & 0.6810  & 0.3140  & 0.1722  & 0.3888  \\ \cdashline{1-6} 
Avg. & 0.3810 & 0.6739  & 0.3093  & 0.1652  & 0.3824  \\ \midrule 

\multicolumn{6}{l}{\textit{GTR-base} (110M) \citep{ni2021large}} \\ \midrule
$k=2$ & 0.3986 & 0.6157  & 0.2930  & 0.1647  & 0.3680  \\  
$k=4$ & 0.3999 & 0.6436  & 0.3065  & 0.1751  & 0.3813  \\  
$k=8$ & 0.4137 & 0.6203  & 0.3096  & 0.1828  & 0.3816  \\ \cdashline{1-6} 
Avg. & 0.4041 & 0.6266  & 0.3031  & 0.1742  & 0.3770  \\ \midrule 

\multicolumn{6}{l}{\textit{GTR-large} (335M) \citep{ni2021large}} \\ \midrule
$k=2$ & 0.4224 & 0.6194  & 0.3405  & 0.1727  & 0.3887  \\  
$k=4$ & 0.4264 & 0.6494  & 0.3652  & 0.1756  & 0.4041  \\  
$k=8$ & 0.4367 & 0.6416  & 0.3693  & 0.1801  & 0.4069  \\ \cdashline{1-6} 
Avg. & 0.4285 & 0.6368  & 0.3583  & 0.1761  & 0.3999  \\ \midrule 

\multicolumn{6}{l}{\textit{GTR-XL} (1.24B) \citep{ni2021large}} \\ \midrule
$k=2$ & 0.4256 & 0.6258  & 0.3852  & 0.1688  & 0.4013  \\  
$k=4$ & 0.4283 & 0.6459  & 0.3902  & 0.1765  & 0.4102  \\  
$k=8$ & 0.4357 & 0.6530  & 0.3902  & 0.1813  & 0.4151  \\ \cdashline{1-6} 
Avg. & 0.4299 & 0.6416  & 0.3885  & 0.1755  & 0.4089  \\ \midrule

\end{tabularx}
}
\end{minipage}
\begin{minipage}{.05\linewidth}
\phantom{This text will be invisible}
\end{minipage}
\begin{minipage}{.475\linewidth}
\centering
\subfloat[nDCG@20 on the test set.]{
\begin{tabularx}{\columnwidth}{@{}lYYYYY@{}}
\toprule
 & Robust & Covid & News & Touche & Avg. \\ \midrule

\multicolumn{6}{l}{\textit{DPR-single}} \\ \midrule
$k=2$ & 0.1064 & 0.4690  & 0.1738  & 0.0824  & 0.2079  \\  
$k=4$ & 0.1144 & 0.4822  & 0.1911  & 0.0815  & 0.2173  \\  
$k=8$ & 0.1230 & 0.5016  & 0.1930  & 0.0849  & 0.2256  \\ \cdashline{1-6} 
Avg. & 0.1146 & 0.4843  & 0.1859  & 0.0830  & 0.2169  \\ \midrule 

\multicolumn{6}{l}{\textit{DPR-multi}} \\ \midrule
$k=2$ & 0.2234 & 0.4622  & 0.2058  & 0.1190  & 0.2526  \\  
$k=4$ & 0.2372 & 0.4558  & 0.2122  & 0.1199  & 0.2563  \\  
$k=8$ & 0.2447 & 0.4846  & 0.2160  & 0.1249  & 0.2675  \\ \cdashline{1-6} 
Avg. & 0.2351 & 0.4675  & 0.2113  & 0.1213  & 0.2588  \\ \midrule 

\multicolumn{6}{l}{\textit{ANCE}} \\ \midrule
$k=2$ & 0.3523 & 0.6826  & 0.3218  & 0.1744  & 0.3828  \\  
$k=4$ & 0.3561 & 0.6887  & 0.3087  & 0.1786  & 0.3830  \\  
$k=8$ & 0.3613 & 0.7023  & 0.3135  & 0.1766  & 0.3884  \\ \cdashline{1-6} 
Avg. & 0.3566 & 0.6912  & 0.3147  & 0.1765  & 0.3847  \\ \midrule 

\multicolumn{6}{l}{\textit{MiniLM}} \\ \midrule
$k=2$ & 0.3531 & 0.6611  & 0.2537  & 0.1637  & 0.3579  \\  
$k=4$ & 0.3652 & 0.6486  & 0.2512  & 0.1649  & 0.3575  \\  
$k=8$ & 0.3677 & 0.6854  & 0.2578  & 0.1687  & 0.3699  \\ \cdashline{1-6} 
Avg. & 0.3620 & 0.6650  & 0.2542  & 0.1658  & 0.3618  \\ \midrule 

\multicolumn{6}{l}{\textit{TAS-B}} \\ \midrule
$k=2$ & 0.3658 & 0.6872  & 0.2813  & 0.1486  & 0.3707  \\  
$k=4$ & 0.3833 & 0.6845  & 0.2830  & 0.1567  & 0.3769  \\  
$k=8$ & 0.3952 & 0.6878  & 0.2787  & 0.1557  & 0.3793  \\ \cdashline{1-6} 
Avg. & 0.3814 & 0.6865  & 0.2810  & 0.1537  & 0.3756  \\ \midrule 

\multicolumn{6}{l}{\textit{GPL}} \\ \midrule
$k=2$ & 0.3623 & 0.6979  & 0.3011  & 0.1617  & 0.3808  \\  
$k=4$ & 0.3800 & 0.7007  & 0.3075  & 0.1663  & 0.3886  \\  
$k=8$ & 0.3873 & 0.7129  & 0.3191  & 0.1681  & 0.3969  \\ \cdashline{1-6} 
Avg. & 0.3766 & 0.7038  & 0.3092  & 0.1654  & 0.3888  \\ \midrule 

\multicolumn{6}{l}{\textit{GTR-base}} \\ \midrule
$k=2$ & 0.3638 & 0.7051  & 0.3177  & 0.1988  & 0.3963  \\  
$k=4$ & 0.3799 & 0.7188  & 0.3262  & 0.2080  & 0.4082  \\  
$k=8$ & 0.3875 & 0.7222  & 0.3250  & 0.2130  & 0.4119  \\ \cdashline{1-6} 
Avg. & 0.3771 & 0.7154  & 0.3230  & 0.2066  & 0.4055  \\ \midrule 

\multicolumn{6}{l}{\textit{GTR-large}} \\ \midrule
$k=2$ & 0.4003 & 0.6844  & 0.3308  & 0.1941  & 0.4024  \\  
$k=4$ & 0.4070 & 0.6851  & 0.3308  & 0.1984  & 0.4053  \\  
$k=8$ & 0.4093 & 0.6850  & 0.3426  & 0.2012  & 0.4095  \\ \cdashline{1-6} 
Avg. & 0.4056 & 0.6848  & 0.3347  & 0.1979  & 0.4057  \\ \midrule 

\multicolumn{6}{l}{\textit{GTR-XL}} \\ \midrule
$k=2$ & 0.4024 & 0.6804  & 0.3456  & 0.2054  & 0.4084  \\  
$k=4$ & 0.4135 & 0.6858  & 0.3387  & 0.2130  & 0.4127  \\  
$k=8$ & 0.4157 & 0.6764  & 0.3441  & 0.2140  & 0.4125  \\ \cdashline{1-6} 
Avg. & 0.4105 & 0.6808  & 0.3428  & 0.2108  & 0.4112  \\ \midrule 

\end{tabularx}
}

\end{minipage}
\caption{Zero-shot nDCG@20 results on the residual collection. Model sizes are reported in parentheses. For DPR, MiniLM, TAS-B and GTR we use the checkpoints from the sentence-transformers library \citep{reimers-gurevych-2019-sentence}. For GPL, we use the self-miner model.}
\label{tab:zero-shot-baselines}
\end{table}

\clearpage 
\twocolumn

\section{BM25-QE and Neural Re-Ranker Top 20 Analysis}\label{sec:appendix-intersection}

\begin{figure}[ht!]
    \centering
    \includegraphics[width=\columnwidth]{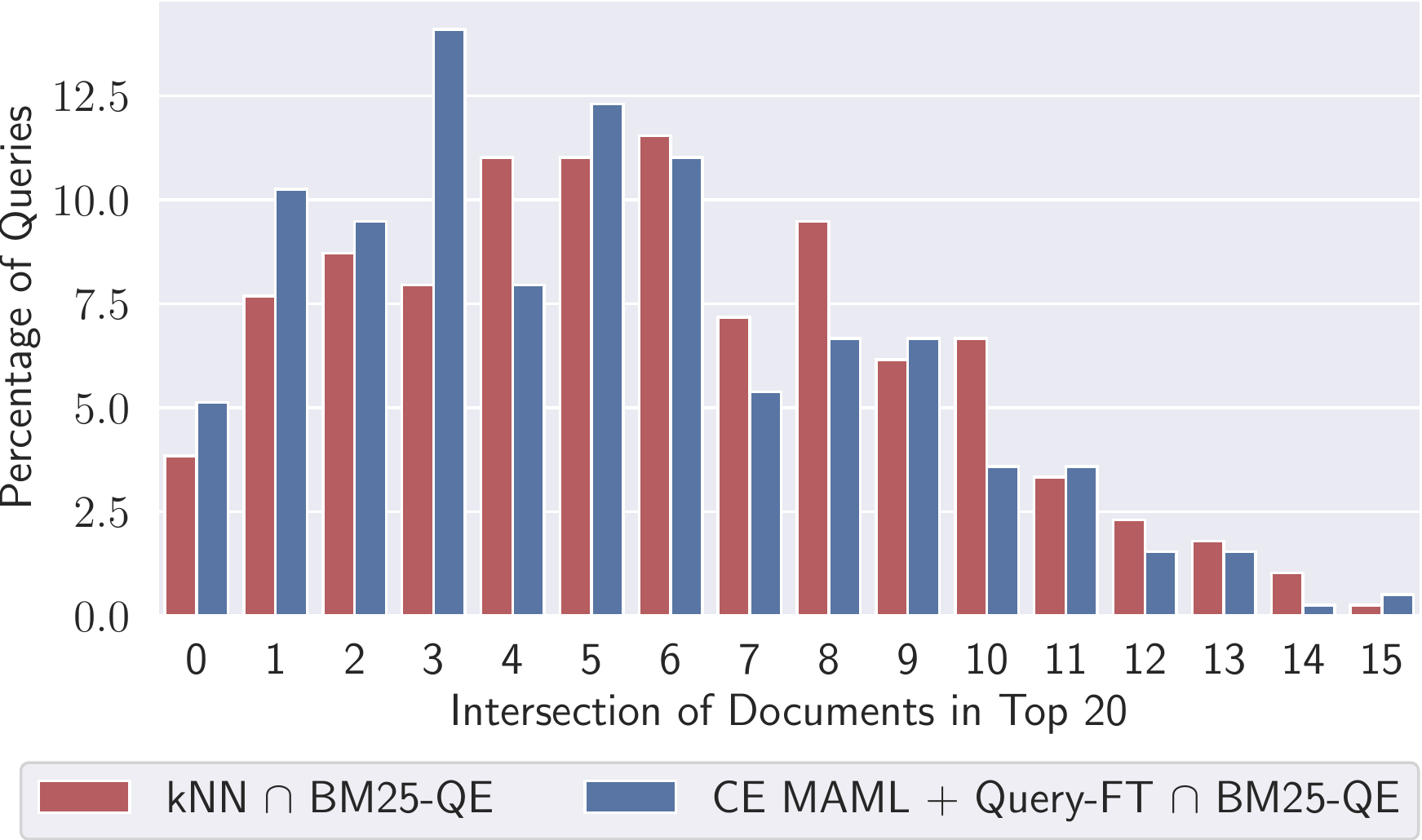}
    \caption{Overlap of documents in the top 20 between BM25-QE and two neural re-ranking methods.}
    \label{fig:overlap}
\end{figure}

Figure~\ref{fig:overlap} presents the number of documents that BM25-QE and the respective neural re-ranking method have in common in the top 20 retrieval results. It shows that while there is some overlap, the methods rank different documents on top.


\section{Retrieval Speed with Query Expansion}\label{sec:appendix-bm25-times}

\begin{figure}[ht!]
\centering
\includegraphics[width=\columnwidth]{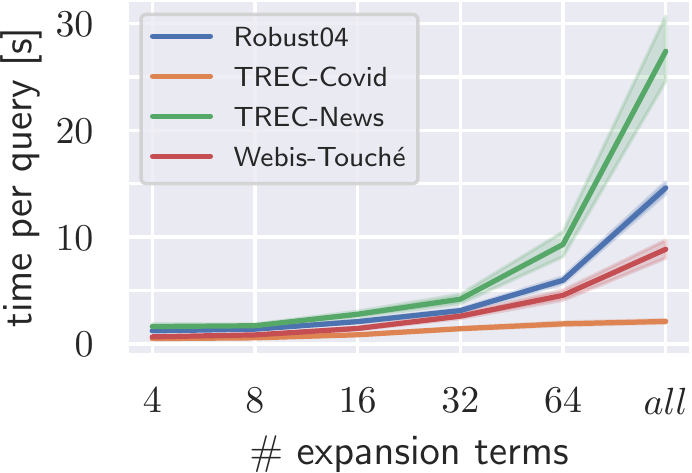}
\caption{Average duration per query with varying the number of expansion terms.}
\label{fig:bm25_times}
\end{figure}

Figure~\ref{fig:bm25_times} presents the average time duration per query when varying the number of expansion terms when using BM25 with query expansion. 

\clearpage
\section{Results Validation Set}
\begin{table}[ht!]
\small
\centering
\begin{tabularx}{\columnwidth}{@{}lYYYYY@{}}
\toprule
 & Robust & Covid & News & Touche & Avg. \\ \midrule
\multicolumn{6}{l}{\textit{BM25-QE}} \\ \midrule
$k=2$ & 0.4135 & 0.6340 & \underline{0.4195} & 0.2113 & 0.4196 \\
$k=4$ & 0.4442 & 0.6103 & 0.4346 & \underline{0.2255} & 0.4287 \\
$k=8$ & 0.4870 & 0.6894 & \underline{0.4798} & \underline{0.2466} & 0.4757 \\ \cdashline{1-6}
Avg. & 0.4483 & 0.6446 & 0.4446 & 0.2278 & 0.4413 \\ \midrule

\multicolumn{6}{l}{\textit{kNN}} \\ \midrule
$k=2$ & 0.4346 & 0.6159 & 0.3702 & 0.1301 & 0.3877 \\
$k=4$ & 0.4267 & 0.6420 & 0.3955 & 0.1523 & 0.4041 \\
$k=8$ & 0.4715 & 0.6771 & 0.4475 & 0.1742 & 0.4426 \\ \cdashline{1-6}
Avg. & 0.4443 & 0.6450 & 0.4044 & 0.1522 & 0.4115 \\ \midrule

\multicolumn{6}{l}{\textit{CE Zero-Shot}} \\ \midrule
$k=2$ & 0.3902 & 0.6598 & 0.2917 & 0.1618 & 0.3759 \\
$k=4$ & 0.3919 & 0.6786 & 0.3249 & 0.1700 & 0.3914 \\
$k=8$ & 0.4064 & 0.6713 & 0.3416 & 0.1799 & 0.3998 \\ \cdashline{1-6}
Avg. & 0.3962 & 0.6699 & 0.3194 & 0.1706 & 0.3890 \\ \midrule

\multicolumn{6}{l}{\textit{CE Query-FT}} \\ \midrule
$k=2$ & 0.4511 & 0.6624 & 0.3179 & 0.1786 & 0.4025 \\
$k=4$ & 0.4676 & 0.7381 & 0.3870 & 0.1913 & 0.4460 \\
$k=8$ & 0.5214 & 0.7551 & 0.4130 & 0.2041 & 0.4734 \\ \cdashline{1-6}
Avg. & 0.4800 & 0.7186 & 0.3726 & 0.1913 & 0.4406 \\ \midrule

\multicolumn{6}{l}{\textit{CE MAML + Query FT}} \\ \midrule
$k=2$ & \underline{0.4640} & 0.6691 & 0.3276 & \underline{0.2186} & 0.4198 \\
$k=4$ & \underline{0.5096} & \underline{0.7595} & 0.3845 & 0.2216 & \underline{0.4688} \\
$k=8$ & \underline{0.5361} & \underline{0.7729} & 0.4035 & 0.2237 & 0.4840 \\ \cdashline{1-6}
Avg. & \underline{0.5032} & \underline{0.7339} & 0.3719 & 0.2213 & 0.4576 \\ \midrule

\multicolumn{6}{l}{\textit{Rank Fusion: kNN \& BM25-QE}} \\ \midrule
$k=2$ & 0.4551 & \underline{0.6992} & \textbf{0.4273} & 0.1931 & \underline{0.4437} \\
$k=4$ & 0.4661 & 0.6877 & \textbf{0.4574} & 0.2144 & 0.4564 \\
$k=8$ & 0.5155 & 0.7261 & \textbf{0.5039} & 0.2243 & \underline{0.4924} \\ \cdashline{1-6}
Avg. & 0.4789 & 0.7043 & \textbf{0.4629} & 0.2106 & \underline{0.4642} \\ \midrule

\multicolumn{6}{l}{\textit{Rank Fusion: CE MAML + Query-FT \& BM25-QE}} \\ \midrule
$k=2$ & \textbf{0.4911} & \textbf{0.7156} & 0.4190 & \underline{0.2389} & \textbf{0.4661} \\
$k=4$ & \textbf{0.5359} & \textbf{0.7610} & \underline{0.4516} & \textbf{0.2468} & \textbf{0.4988} \\
$k=8$ & \textbf{0.5724} & \textbf{0.7998} & 0.4708 & \textbf{0.2475} & \textbf{0.5226} \\ \cdashline{1-6}
Avg. & \textbf{0.5331} & \textbf{0.7588} & \underline{0.4471} & \textbf{0.2444} & \textbf{0.4958} \\ \bottomrule
\end{tabularx}
\label{tab:main-valid}
\caption{nDCG@20 results on the validation set. The top-performing result is shown in boldface, runner-up is underlined.}
\end{table}

\vfill\eject 

\section{Ablations Validation Set}
\begin{table}[ht!]
\small
\centering
\begin{tabularx}{\columnwidth}{@{}lYYYYY@{}}
\toprule
 & Robust & Covid & News & Touche & Avg. \\ \midrule
\multicolumn{6}{l}{\textit{BM25 without feedback documents}} \\ \midrule
& 0.0451 & 0.1802 & 0.0403 & 0.0861 & 0.0879 \\ \midrule

\multicolumn{6}{l}{\textit{kNN (Query Only)}} \\ \midrule
$k=2$ & 0.3824 & 0.5801 & 0.2800 & 0.1433 & 0.3464 \\
$k=4$ & 0.3843 & 0.5847 & 0.2831 & 0.1570 & 0.3523 \\
$k=8$ & 0.4044 & 0.6265 & 0.2856 & 0.1668 & 0.3708 \\ \cdashline{1-6}
Avg. & 0.3903 & 0.5971 & 0.2829 & 0.1557 & 0.3565 \\ \midrule

\multicolumn{6}{l}{\textit{CE Query-FT (full)}} \\ \midrule
$k=2$ & 0.4876 & 0.6808 & 0.3397 & 0.2005 & 0.4272 \\
$k=4$ & 0.5003 & 0.7700 & 0.3896 & 0.2129 & 0.4682 \\
$k=8$ & 0.5598 & 0.7802 & 0.4180 & 0.2074 & 0.4914 \\ \cdashline{1-6}
Avg. & 0.5159 & 0.7437 & 0.3824 & 0.2069 & 0.4622 \\ \midrule

\multicolumn{6}{l}{\textit{CE supervised + Query-FT (bias)}} \\ \midrule
$k=2$ & 0.4579 & 0.7260 & 0.3189 & 0.2131 & 0.4290 \\
$k=4$ & 0.4876 & 0.7286 & 0.3854 & 0.2256 & 0.4568 \\
$k=8$ & 0.5200 & 0.7449 & 0.4241 & 0.2296 & 0.4797 \\ \cdashline{1-6}
Avg. & 0.4885 & 0.7332 & 0.3761 & 0.2228 & 0.4551 \\ \midrule
\end{tabularx}
\label{tab:ablations-valid}
\caption{nDCG@20 results on the validation set of the ablation studies.}
\end{table}

\clearpage

\onecolumn
\section{Models, Hyperparameters \& Hardware}
\begin{table}[!ht]
\centering
\begin{tabular}{ll}
\toprule
\multicolumn{2}{l}{\textit{Hardware \& Settings for Latency Experiments}} \\ \midrule
Elasticsearch Version & 7.11.2 \\
Elasticsearch Settings & Single Shard, Cache cleared after each query \\
Elasticsearch Hardware & 24 Intel Xeon CPU E5-2620 v2 @ 2.10GHz \\
GPU (for kNN and CE) & NVIDIA P100, 16GB \\
\midrule
\multicolumn{2}{l}{\textit{Models}} \\ \midrule
kNN & \href{https://huggingface.co/sentence-transformers/all-MiniLM-L6-v2}{sentence-transformers/all-MiniLM-L6-v2}\\ 
Cross-Encoder & \href{https://huggingface.co/cross-encoder/ms-marco-MiniLM-L-6-v2}{cross-encoder/ms-marco-MiniLM-L-6-v2} \\\midrule
\multicolumn{2}{l}{\textit{Model Parameters}} \\ \midrule
kNN & 22.7M \\
Cross-Encoder & 22.7M \\
Cross-Encoder (biases only) & 26k \\ \midrule
\multicolumn{2}{l}{\textit{Training Settings}} \\ \midrule
Optimizer & AdamW \citep{adamW} \\
Optimizer (MAML) & SGD \\ \midrule
\multicolumn{2}{l}{\textit{Hyperparameters}} \\ \midrule
\begin{tabular}[c]{@{}l@{}}Learning rates for Query-FT,\\MAML and supervised training\end{tabular} & $\{\num{2e-3}, \num{2e-4}, \num{2e-5}\}$ \\
Epochs query fine-tuning & $1-8$ \\ \midrule
\multicolumn{2}{l}{\textit{Evaluation Libraries}} \\ \midrule
\textsc{pytrec-eval} Version & 0.5 \\
\bottomrule
\end{tabular}
\caption{Hyperparameters and models used in our experiments. The best learning rate and the number of epochs have been selected according to the nDCG@20 validation performance.}
\label{tab:my_label}
\end{table}
\section{2\textsuperscript{nd} Stage Retrieval: BM25 with Query Expansion}\label{sec:bm25-expansion}

\begin{table}[h!]
\begin{minipage}{.45\linewidth}
\centering
\subfloat[Recall@1000 on the validation set.]{
\begin{tabularx}{\columnwidth}{@{}lYYYY@{}}
\toprule
$\mathtt{e}$ & $\mathtt{k}=2$ & $\mathtt{k}=4$ & $\mathtt{k}=8$ & Avg. \\ \midrule
4 & 0.6164 & 0.6242 & 0.6520 & 0.6309 \\
8 & \underline{0.6167} & \underline{0.6369} & \underline{0.6686} & \underline{0.6407} \\
16 & \textbf{0.6266} & \textbf{0.6470} & \textbf{0.6700} & \textbf{0.6479} \\
32 & 0.6122 & 0.6262 & 0.6463 & 0.6283 \\
64 & 0.5704 & 0.5776 & 0.5868 & 0.5783 \\
all & 0.5722 & 0.5828 & 0.5823 & 0.5791 \\ \bottomrule
\end{tabularx}
}
\end{minipage}
\begin{minipage}{.1\linewidth}
\phantom{This text will be invisible}
\end{minipage}
\begin{minipage}{.45\linewidth}
\centering
\subfloat[Recall@1000 on the test set.]{
\begin{tabularx}{\columnwidth}{@{}lYYYY@{}}
\toprule
$e$ & $k=2$ & $k=4$ & $k=8$ & Avg. \\ \midrule
4 & 0.6187 & 0.6300 & 0.6566 & 0.6351 \\
8 & \textbf{0.6280} & \textbf{0.6414} & \underline{0.6721} & \textbf{0.6472} \\
16 & \underline{0.6195} & \underline{0.6400} & \textbf{0.6736} & \underline{0.6444} \\
32 & 0.6039 & 0.6209 & 0.6477 & 0.6242 \\
64 & 0.5597 & 0.5729 & 0.5843 & 0.5723 \\
all & 0.5723 & 0.5771 & 0.5828 & 0.5774 \\ \bottomrule
\end{tabularx}
}
\end{minipage}
\caption{Recall@1000 results for BM25 with query expansion on the validation (a) and test (b) set for varying number of expansion terms $e$ extracted per document. In bold best performing setting and in underline runner-up. The last column shows the average over all $k$. Although using $e = 8$ performs best on the test set, we conduct subsequent experiments with $e = 16$ since we also tune other hyperparameters on the validation set.}
\label{tab:bm25-expansion}
\end{table}

\end{document}